\documentclass[%
	a4paper,
	twoside,
	bibliography=totoc,
	listof=totoc,
	index=totoc,
	parskip=half,
	chapterprefix=false,
	numbers=noenddot,
	11pt
]{article} 

\usepackage{authblk}
\usepackage[left=3cm,right=2.5cm,top=3cm,bottom=4cm]{geometry}  
\usepackage{rotating}                                           
\usepackage[singlespacing]{setspace}                           
\usepackage{indentfirst}                                        
\usepackage{titlesec}                                           
\titlespacing{\section}{0pt}{36pt}{6pt}                         
\titlespacing{\subsection}{0pt}{24pt}{6pt}                      
\usepackage[toc,page]{appendix}                                 
\usepackage{parskip}

\usepackage{color}						
\usepackage{ulem}						

\usepackage{scrpage2}					        
\clearscrheadfoot                                               
\setheadsepline{1pt}                                            
\setfootsepline{1pt}                                            
\lohead[]{\leftmark}                                            
\rehead[]{\rightmark}                                           

\usepackage{array}						
\usepackage{booktabs}                   			
\usepackage{longtable}						
\usepackage{rotating}                                           
\usepackage{multirow}						

\usepackage{graphicx}						
\usepackage{subfig}                     			
\usepackage{caption}						
\usepackage{subfig}                                             

\usepackage{amsfonts}						
\usepackage{amssymb}						
\usepackage{amsmath}						
\usepackage{gensymb}

\usepackage[square,numbers]{natbib}                                 
\usepackage{acronym}                                            

\begin{document}

\newcommand{\HRule}{\rule{\linewidth}{0.5mm}}
\newcommand\redsout{\bgroup\markoverwith{\textcolor{black}{\rule[0.5ex]{2pt}{0.8pt}}}\ULon}

\title{Convergence of extreme value statistics in a two-layer quasi-geostrophic atmospheric model}
\author{Vera Melinda G\'{a}lfi ($1$,$2$), Tam\'{a}s B\'{o}dai ($3$), Valerio Lucarini ($1,3,4$)}

\affil{(1) Meteorological Institute, CEN, University of Hamburg, Hamburg, Germany, (2) IMPRS-ESM, Max Planck Institute for Meteorology, Hamburg, Germany, (3) Department of Mathematics and Statistics, University of Reading, Reading, UK, (4) Center for Environmental Policy, Imperial College London, London, UK}

\maketitle

\pagenumbering{arabic}
\setcounter{page}{1}
\cfoot[\pagemark]{\pagemark}                                    

\begin{abstract}
We search for the signature of universal properties of extreme events, theoretically predicted for Axiom A flows, in a chaotic and high dimensional dynamical system by studying the convergence of GEV (Generalized Extreme Value) and GP (Generalized Pareto) shape parameter estimates to a theoretical value, expressed in terms of partial dimensions of the attractor, which are global properties. We consider a two layer quasi-geostrophic (QG) atmospheric model using two forcing levels, and analyse extremes of different types of physical observables (local, zonally-averaged energy, and the average value of energy over the mid-latitudes). Regarding the predicted universality, we find closer agreement in the shape parameter estimates only in the case of strong forcing, producing a highly chaotic behaviour, for some observables (the local energy at every latitude). Due to the limited (though very large) data size and the presence of serial correlations, it is difficult to obtain robust statistics of extremes in case of the other observables. In the case of weak forcing, inducing a less pronounced chaotic flow with regime behaviour, we find worse agreement with the theory developed for Axiom A flows, which is unsurprising considering the properties of the system. 
\end{abstract}

\section{Introduction and motivation}
\label{sec:intro}

The investigation of extreme events is extremely relevant for a range of disciplines in mathematical, natural, social sciences and engineering. Understanding the large fluctuations of the system of interest is of great importance from a theoretical point of view, but also when it comes to assess the risk associated to low probability and high impact events. In many cases, in order to gauge preparedness and resilience properly, one would like to be able to quantify the return times for events of different intensity, and take suitable measures for preventing the expected impacts. Prominent examples are weather and climate extremes, which can have a huge impact on human society and natural ecosystems. The present uncertainty in the future projections of extremes makes their study even more urgent and crucial \citep{ipcc2013}.

In practical terms, the main goal behind the study of the extreme events is to understand the properties of the highest quantiles of the variable of interest. A fundamental drawback comes from the fact that extreme events are rare, so that it is difficult to collect satisfactory statistics from the analysis of a time series of finite length. Additionally, in the absence of a strong mathematical framework, it is virtually impossible to make quantitative statements about the probability of occurrence of events larger than observed. Therefore, statistical inference based on empirical models tends to suffer from the lack of predictive power.
A robust theoretical framework for analysing extreme events is provided by extreme value theory (EVT). After the early contributions by \citet{fisher1928}, EVT was introduced by \citet{gnedenko1943}, who discovered that under rather general conditions the extreme events associated to stochastic variables can be described by a family of parametric distributions. As a result, the problem of studying the extremes of a stochastic variable can be reduced to estimating the parameters of a known probability distribution.  The most important parameter of such a distribution is called the shape parameter, which determines the qualitative properties of the distribution, and, in particular, whether it has a finite or infinite upper point, or, more concretely, whether extremes are bounded by an absolute finite maximum or not. 

The two most important properties of extremes, i.e., that they occur rarely and their magnitude is unusually high or low, constitute the basis of two popular methods of EVT, the block maxima (BM) and the peak-over-threshold (POT) approaches. The BM approach aims at finding the limiting distribution of maxima $M_n$ of independent identically distributed random (i.i.d.r.) variables separated into blocks of size $n$, as $n \to \infty$. Under rather general conditions and for many different parent distributions, the limiting distributions of block maxima belong to the Generalized Extreme Value (GEV) distribution family; this is the point of view originally proposed by \citet{gnedenko1943}. The POT approach aims at understanding the statistical properties of the exceedances of a stochastic variable $X$ above a given threshold $u$. Under the same conditions as for the BM approach, if $u$ is very high, the distribution of the threshold exceedances $X-u$ belongs to the Generalized Pareto (GP) distribution family \citep{pickands1975, balkemadehaan1974}. The existence of well-defined functional forms for the distributions describing extreme events provides predictive power: one can in principle compute the return time of some yet unobserved events. It is remarkable that these two points of view on extremes (which lead to different selection procedures and different choice of the events classified as extremes) are, in fact, equivalent. 

Problems in applying EVT to actual time series result from the fact that, typically, the observed data feature a certain degree of serial correlations \citep{ghiletal2011}. Note that the setting behind the construction of EVT can be extended by relaxing the hypothesis of independence of the random variables, as long as correlations are decaying sufficiently fast. This is of clear relevance when trying to use EVT for studying observables of deterministic dynamical systems. In this case, in fact, the underlying dynamics determines the existence of correlations between the values of observables at different times, and one can easily guess that, when the dynamical system is chaotic, there is good hope of deriving EVT for its observables \citep{lucarini2016}. Obtaining the true limiting EVT can be extremely hard, even in simple dynamical systems \citep{faranda2011}. When analysing finite time series, the convergence of the estimated GEV or GP shape parameters to the asymptotic values can be very slow. The speed of convergence depends on the type of parent distribution \citep{leadbetteretal1983}, and can be additionally slowed down by correlations \citep{rust2009, coles2001}. Due to the fact that the data size is always limited, there is typically a difference between the asymptotic GEV or GP parameters and the estimated ones; finite-size estimates are generally biased. For example, the GEV shape parameter of a simple Gaussian process is 0, but, for any finite time series, we would estimate typically a negative shape parameter \citep{fisher1928}.

When performing statistical inference using the BM or POT method (fitting the GEV or GP model, respectively, to data), it is crucial to have an appropriate protocol of selection of ``good" candidates for extremes \citep{coles2001}. On the one hand, if the chosen blocks (for the BM method) are too short or the threshold (for the POT method) is too low, the approximation of the limit model is likely to be inappropriate, leading to false parameter estimates. Hence, the verification of the agreement between the statistical model and the available data is essential, which is often done based on goodness-of-fit tests, like the Kolmogorov-Smirnov \citep{massey1951}, Anderson-Darling \citep{andersondarling1954} or Pearson's chi-squared tests \citep{agresti2007}. On the other hand, if the blocks are too large or the threshold is too high, the number of extremes may be insufficient for a reliable estimation of the parameters, and uncertainty becomes very high. As discussed later in the paper, \citet{coles2001} shows how to derive an optimal choice for the value of the block size or the threshold, in such a way as to verify that we are close to the asymptotic level as required by EVT but we use the available data as efficiently as possible.

Classical EVT has been extended and adapted to analyse extremes of observables of chaotic dynamical systems, where the sensitive dependence on initial conditions is fundamentally responsible for generating a de-facto stochastic process. The reader is referred to \citet{lucarini2016} for a detailed overview of the field of EVT for dynamical systems. It is possible to establish an EVT for observables of chaotic dynamical systems when one considers Axiom A dynamical systems. These are rather special chaotic dynamical systems that are uniformly hyperbolic on their attractor (so that stable and unstable directions are well separated), which supports a so-called Sinai-Ruelle-Bowen (SRB) measure. Such an invariant measure has physical relevance because it is stable against weak stochastic perturbations \citep{eckmannruelle1985, ruelle1989}. One of the great merits of Axiom A dynamical systems is that they allow for deriving rigorous and robust statistical mechanical properties for purely deterministic background dynamics. Despite having deterministic dynamics, when looking at their observables, they behave just like generators of stochastic processes. While Axiom A systems are rather special and indeed not generic, they have great relevance for applications if one takes into account the chaotic hypothesis, which indicates that high-dimensional chaotic systems behave at all practical purposes as if they were Axiom A \citep{gallavotticohen1995, gallavotti2014}. 

Several studies dealing with EVT for dynamical systems reveal a link between the statistical properties of the extremes and geometric (and possibly in turn global dynamical) characteristics of the system producing these extremes \citep{faranda2011, lucarini2012a, lucarini2012b, holland2012, lucarini2014}. The main findings are that when suitable observables are chosen for the dynamical system of interest, it is possible to relate the GEV or GP parameters describing the extremes to basic properties of the dynamics, and especially to the geometry of the attractor. In particular, depending on the choice of the observable, one can associate the most important parameter of the GEV or GP distribution to the information dimension of the attractor or to the partial information dimension along the stable and unstable directions of the flow~\citep{lucarini2014}. These partial dimensions are well-defined everywhere on the chaotic attractor, possibly with a variation with location, also for non-uniformly hyperbolic systems~\citep{10.2307/121072}, beside Axiom A systems. However, Axiom A systems possess an ergodic SRB measure which lends itself to a universality of the shape parameter for all sufficiently smooth observables; the local or point-wise (partial) dimensions taking the same value almost everywhere~\citep{ott1993}. In this case the uniform shape parameter can be related to the (partial) Kaplan-Yorke dimension(s) which is (are) defined by the global dynamical characteristic numbers, the Lyapunov exponents. Clearly, this is an asymptotic result, and one must expect that differences emerge on pre-asymptotic level when different observables are studied. In other words, this theory does not make predictions regarding the convergence of shape parameter estimates, the analysis of which is the main objective of this paper. Via the connection with fractal dimensions, it can be said that the analysis of extremes acts as a microscope able to assess the fine scale properties of the invariant measures.

Some preliminary numerical tests show that the convergence to the asymptotic shape parameter is slow in low-dimensional cases, in systems well known not to be Axiom A \citep{lucarini2014}. \citet{bodai2017} examined the convergence to the GEV distribution in the case of extremes of site variables in the Lorenz 96 model \citep{lorenz1996}, investigating separately a range of cases extending from weak to strong chaos. He found that when considering configurations supporting weak chaos with a low-dimensional attractor, the theoretical results obtained in the context of the Axiom A hypothesis are hard to verify. For lower dimensions, up to a dimension of about 5, shape parameter estimates fluctuate greatly rather than converge, while block maxima data can be shown not to conform to a GEV model; and for somewhat larger dimensions, up to 9 in the study, estimates could diverge from the predicted value while data already conform to a GEV model. Good agreement with the theory was found only in the highly turbulent case possessing a higher-dimensional attractor, about 30, supporting the basic idea behind the chaotic hypothesis. Also in this case, nonetheless, very slow convergence was found.

In previous analysis performed on higher dimensional, intermediate complexity models with O($10^2-10^3$) degrees of freedom, very slow (if any) convergence to EVT distributions could be found in the case of extremes of local temperature observables \citep{vannitsem2007}. In another analysis of a similar model~\citep{felici2007a}, the agreement of the distribution of global energy extremes with a member of the GEV family was indeed good, yet large uncertainty remained on the value of the shape parameter, and no stringent test was made to make sure that the estimate was stable against changes in the block size considered in the BM analysis. Clearly, the specific choice of the observable and the degree of chaoticity of the underlying dynamics is of primary relevance regarding the convergence to the limiting GEV or GP distribution.

In this work, we use a quasi-geostrophic (QG) atmospheric model of intermediate complexity featuring 1056 degrees of freedom, to analyse extremes of different types of observables: local energy (defined at each grid point), zonally-averaged energy, and the average value of energy over the mid-latitudes. Our main objective is to compare the estimated GEV and GP shape parameters with a shape parameter derived, based on the theory referred to above, from the properties of the attractor along the stable, unstable, and neutral directions. We refer to this as the ``theoretical shape parameter''. Thus we explore numerically the link between the purely statistical properties of extreme events based on EVT and the dynamical properties of the system producing these extremes. We perform simulations applying two different levels of forcing: a strong forcing, producing a highly chaotic behaviour of the system, and a weak forcing, producing a less pronounced chaotic behaviour. The dimensionality of the attractor is much larger in the former than in the latter case. This work goes beyond the previously mentioned studies based on more simple dynamical systems, in a sense that with our model we can study the convergence for observables being different physical quantities, or, representing different spatial scales/characteristics of the same physical quantity. Additionally, compared to previous studies also performed on intermediate complexity models, we consider longer time series and a variety of observables. Our model is simple compared to a GCM (General Circulation Model), but contains two of the main processes relevant for mid-latitude atmospheric dynamics, i.e., baroclinic and barotropic instabilities. Hence, we contribute to bridging the gap between the analysis of extremes in simple and very high dimensional dynamical systems, as in the case of the GCMs used for atmospheric and climate simulations, by using a model that simulates to a certain degree Earth-like atmospheric processes and allows also for computing with feasible computational costs some dynamical system properties, like Lyapunov Exponents or Kaplan-Yorke Dimensions. The properties of the model have been extensively studied by \citet{schubert2015, schubertlucarini2016}. 

Based on numerical results (Sebastian Schubert, personal communication), the model is expected to be non-hyperbolic, but one can assume that the chaotic hypothesis applies to it (in case of sufficiently high forcing levels inducing a chaotic behaviour of the system), and so in analysing the convergence of shape parameter estimates one can take the predicted theoretical shape parameter as reference. We also assume that the symmetry of the model with respect to longitude, that introduces a central direction besides the directions of expansion and contraction in phase space, does not alter the ergodicity of the system at a practical level, and hence the true shape parameter remains uniform.

Although we use an idealised model, our results are transferable to time series obtained from more realistic model simulations or from measurements. By understanding the differences among the analysed observables, we gain insight into the statistical properties of extremes of geophysical observables with different spatial scales. By using two forcings, we are able to study the convergence to theoretical shape parameters related to different chaotic systems: one exhibiting fast decaying correlations and another one characterised by slower decaying correlations. These aspects are relevant in the case of geophysical applications where one deals also with time series on several spatial scales and with different degrees of correlations.

The structure of this article is as follows. Sec. 2 gives a theoretical overview, describing the block maxima and the peak over threshold approaches. In Sec. 3 we present our model, the performed simulations and the applied methods. In Sec. 4 we discuss our results regarding the statistics of extremes for strong forcing and for weak forcing. We summarise and discuss our results in Sec. 5.

\section{Some Elements of Extreme Value Theory}
\label{sec:theory}

Let us consider \(M_{n} = \mathrm{max}\{X_{1},...,X_{n}\}\), where \(X_{1},...,X_{n} \) is a sequence of i.i.d.r. variables with common distribution function \(F(x)\). The extremal types theorem \citep{fisher1928,gnedenko1943} states that if there exist sequences of constants \(\{a_{n} > 0\}\) and \(\{b_{n} \}\), so that the distribution of normalised $M_n$, i.e., $\mathrm{Pr}\{(M_{n}-b_{n})/a_{n}\le z\}$, converges for \(n \to \infty \) to a non-degenerate distribution function $G(z)$, then $G(z)$ is one of three possible types of so-called extreme value distributions, having the cumulative distribution function
\begin{equation}
  G(z) = 
  \begin{cases}
     \exp\left\{-\left[1+\xi\left(\frac{z-\mu}{\sigma}\right)\right]^{-1/\xi}\right\} & \quad \mathrm{for} \ \xi \ne 0,\\
     \exp\left\{-\exp\left[-\left(\frac{z-\mu}{\sigma}\right)\right]\right\}          & \quad \mathrm{for} \ \xi = 0,\\
  \end{cases}
  \label{eq:GEV1}
\end{equation} 
where \(-\infty<\mu<\infty\), \(\sigma>0\), $1+\xi(z-\mu)/\sigma>0$ for $\xi \ne 0$ and $-\infty<z<\infty$ for $\xi=0$ \citep{coles2001}.

\(G(z)\) represents the GEV family of distributions with three parameters: the location parameter \(\mu\), scale parameter \(\sigma\), and shape parameter \(\xi\). The shape parameter \(\xi\) describes the tail behaviour, and determines to which one of the three types of extreme value distributions \(G(z)\) belongs. If \(\xi=0\), the tail decays exponentially, \(G(z)\) is a type I extreme value distribution or Gumbel distribution. If \(\xi>0\), the tail decays polynomially, and \(G(z)\) belongs to the type II or Fr\'{e}chet distribution. If \(\xi<0\), the domain of the distribution has an upper limit, and is referred to as a type III or Weibull distribution.

Under the same conditions, for which the distribution of $M_{n}$ converges to the GEV distribution, the exceedances \(y=X-u\) of a threshold $u$ reaching the upper right point of the distributions of $X$, given that $X>u$, are asymptotically distributed according to the Generalized Pareto (GP) distribution family \citep{coles2001}
\begin{equation}
  H(y) = 
  \begin{cases}
    1-\left(1+\frac{\tilde\xi y}{\tilde{\sigma}}\right)^{-1/\tilde\xi}     & \quad \mathrm{for} \ \tilde\xi \ne 0,\\
    1-\exp\left(-\frac{y}{\tilde{\sigma}}\right)               & \quad \mathrm{for} \ \tilde\xi = 0,\\
  \end{cases}
  \label{eq:GPD1}
\end{equation}
where $1+\tilde\xi y/\tilde{\sigma}>0$ for $\tilde\xi \ne 0$, $y>0$, and $\tilde{\sigma}>0$. $H(y)$ has two parameters: the scale parameter $\tilde{\sigma}$ and the shape parameter $\tilde\xi$. The shape parameter \(\tilde\xi\) describes again the tail behaviour, and determines to which one of the three types of GP distributions $H(y)$ belongs. If $\tilde\xi=0$, the tail of the distribution decays exponentially; if \(\tilde\xi>0\), the tail decays polynomially; and if \(\tilde\xi<0\) the distribution is bounded \citep{pickands1975,balkemadehaan1974,davison1990}. If convergence to the GEV and GP distributions is realised, $\tilde\xi=\xi$ and $\tilde\sigma=\sigma+\xi(u-\mu)$. As a result, once we estimate the parameters for the GEV, we can derive the corresponding GP parameters, and vice versa \citep{coles2001}. 

From the values of the GEV or GP parameters it is possible to infer the expected return levels or extreme quantiles. Return levels $z_p$ are obtained from the GEV distribution by inverting equation (\ref{eq:GEV1}):
\begin{equation}
  z_p =
  \begin{cases}
    \mu-\frac{\sigma}{\xi}\left[1-y_{p}^{-\xi}\right]       & \quad \mathrm{for} \ \xi \neq 0,\\
    \mu-\sigma \ \mathrm{log} \ y_p                         & \quad \mathrm{for} \ \xi = 0,\\
  \end{cases}
  \label{eq:rlgev}
\end{equation}
where $y_p=-\log({1-G(z_p)})$, and $1/y_p$ represents the return period. In the case of the GP distribution, the $m$-observation return level $z_m$ (i.e., the level that is exceeded on average every $m$ observations) can be derived from equation (\ref{eq:GPD1}):
\begin{equation}
  z_m=
  \begin{cases}
    u-\frac{\tilde{\sigma}}{\tilde\xi}\left[1-(\frac{1}{m \zeta_u})^{-\tilde\xi}\right]       & \quad \mathrm{for } \ \tilde\xi \neq 0,\\
    u-\tilde{\sigma} \mathrm{ log } (\frac{1}{m \zeta_u})                    & \quad \mathrm{for } \ \tilde\xi = 0,\\
  \end{cases}
  \label{eq:rlgpd}
\end{equation}
where $\zeta_u$ represents the probability of an individual observation exceeding the threshold $u$ \citep{coles2001}. By plotting the GEV (GP) return level $z_p$ ($z_m$) against the return period $1/y_p$ ($m\zeta_u$) on a logarithmic scale, the plot is linear if $\xi=0$ ($\tilde\xi=0$), is convex if $\xi>0$ ($\tilde\xi>0$), and is concave if $\xi<0$ ($\tilde\xi<0$).

In the case of a correlated stationary stochastic process the same GEV limit laws apply as for i.i.d.r. variables if certain conditions, regarding the decay of serial correlation, are fulfilled \citep{leadbetter1974, leadbetter1989,lucarini2016}. By stationary, we refer to a sequence of correlated variables whose joint probability distribution is time-invariant. \textcolor{black}{However, an important restriction is that, as an effect of serial correlation, an effective block size can be defined which is smaller than the number of observations in a block}. This can enhance the bias in the parameter estimation, appearing as a slower or delayed convergence of the block maxima distribution to the limiting GEV distribution \citep{coles2001, rust2009}. Another possible effect of serial correlation is the appearance of extremes at consecutive time steps (clusters). If an extreme value law does exist in this case, then $G^*(z)=G(z)^\theta$, where $\theta$ is called the extremal index and $0<\theta<1$ ($G^*(z)$ denotes the 
limiting distribution of BM from the correlated sequence and $G(z)$ the one from an uncorrelated sequence, having the same marginal distribution). Clusters of extremes represent a problem especially when applying the POT approach. A widely-adopted method to get rid of the correlated extremes is declustering, which basically consists of identifying the maximum excess within each cluster, and fitting the GP distribution to the cluster maxima \citep{leadbetter1989,smith1989,ferro2003}.

As mentioned before, several studies on EVT for observables of dynamical systems relate the GEV and GP shape parameters to certain properties of the attractor. In the case of the so-called ``distance'' observables, one can relate the GEV and GP parameters to basic geometrical properties of the attractor \citep{faranda2011, lucarini2012a, lucarini2012b}. The distance observables $g(\text{dist}(x,x_0))$ are functions of the Euclidean distance between one point on the attractor $x_0$ and the orbit $x$. The function $g(y)$ is chosen in a way to have a global maximum for $y=0$, so that large values of $g$ correspond to recurrences of the orbit near $x_0$. Depending on the choice of the function $g(y)$, the extremes of the distance observables can have positive, negative, or vanishing value for the shape parameter. In particular, when $g(y)$ is chosen to be a power law, the shape parameter is non-zero, and it is proportional to the inverse of the Kaplan-Yorke dimension of the attractor; see below for further details \citep{lucarini2012a, lucarini2012b, lucarini2014}.

While recurrence properties are indeed important for characterising a system, distance observables are not well suited for studying some basic physical properties, such as, in the the case of fluids, energy or enstrophy. Hence, \citet{holland2012} studied the extremes of smooth functions $A=A(x)$ which take their maximum on the attractor in a point where the corresponding level surface of $A(x)$ is tangential to the unstable manifold of the attractor, referring to them as ``physical'' observables. They found a relationship between the GEV shape parameter and some geometrical properties of the attractor dealing with the properties of the unstable and stable directions in the tangent space. The results of \citet{holland2012} were re-examined by \citet{lucarini2014}, using the POT approach for physical observables of Axiom A systems. They considered the time-continuous time series of physical observables, and found that for all non-pathological physical observables $A$ the shape parameter can be written as:
\begin{equation}
 \xi_\delta=-\frac{1}{\delta},
 \label{eq:xi} 
\end{equation}
with $\delta$ defined as
\begin{equation}
 \delta=d_s+(d_u+d_n)/2,
 \label{eq:delta} 
\end{equation}
where $d_s$, $d_u$, and $d_n$ are the partial dimensions of the attractor restricted to the stable, unstable, and neutral (i.e., central) directions. \textcolor{black}{As mentioned in Sec.~\ref{sec:intro}, these local or point-wise (partial) dimensions take the same value almost everywhere on the attractor if one considers smooth observables of Axiom A systems.} $d_u$ is equal to the number of positive Lyapunov exponents \citep{ott1993}, $d_n$ is equal to the number of zero Lyapunov exponents, and $d_s=d_{KY}-d_u-d_n$, where $d_{KY}=n+\frac{\sum_{k=1}^n \lambda_k}{|\lambda_{n+1}|}$ is the Kaplan-Yorke dimension with $\lambda_k$ denoting the Lyapunov exponents of the system, in a descending order, and $n$ is such that $\sum_{k=1}^n \lambda_k$ is positive and $\sum_{k=1}^{n+1} \lambda_k$ is negative. We remark that a more general point of view, taking into consideration possible geometrical degeneracies, suggests that $-1/\xi_\delta < d_{KY} < -2/\xi_\delta$, and, additionally, $d_{KY}/2=(d_s+d_u+d_n)/2 \leq \delta \leq d_s+(d_u+d_n)/2$ \citep{lucarini2016}.

According to Eq. (\ref{eq:xi}) the shape parameter is always negative (due to the compactness of the attractor), and it is close to zero in the case of systems having large values of the Kaplan-Yorke dimension. Furthermore, it shows a universal property of extremes, which does not depend on the chosen observable but only on the geometry of the attractor. In what follows we will focus on comparing Eq. (\ref{eq:xi}) with statistically inferred GEV and GP shape parameters in the case of energy extremes of the model of interest in this study, which is described next.

\section{Model description and methods}

We consider a spectral quasi-geostrophic (QG) 2-layer atmospheric model similar to the one introduced by \citet{phillips1956}. Specifically, our model, including the simulation code, is the same as in \citet{schubert2015}, and is a modified version of the one presented by \citet{frisius1998}. The model represents synoptic scale mid-latitude atmospheric dynamics based on the quasi-geostrophic approximation, which assumes hydrostatic balance and allows only small departures from the geostrophic balance. The model features baroclinic conversion and barotropic stabilisation processes, and simulates a turbulent jet-like zonal flow when suitable values are chosen for the parameters of the system. The reader is referred to \citet{holton2004} for a detailed physical and mathematical description of quasi-geostrophic approximation for mid-latitude atmospheric dynamics.
\subsection{Model description}

The model domain is a rectangular channel with latitudinal and longitudinal coordinates $(x,y) \in [0\ L_x] \times [0\ L_y]$. $y=0$ represents the Equator, and $y=L_{y}$ corresponds to the north Pole. We assume periodicity along the $x$-direction, so that $L_x$ corresponds to the length of the parallel at $45\degree$ N. The vertical structure of the model atmosphere consists of only two discrete layers: this is the minimal vertical resolution needed to represent baroclinic processes \citep{holton2004}. Five vertical pressure levels define the two layers with boundaries at: $p_{2.5}=1000$ hPa (surface level), $p_2=750$ hPa, $p_{1.5}=500$ hPa, $p_1=250$ hPa, $p_{0.5}=0$ hPa (top level). The geostrophic stream function $\psi$ is defined at levels $p_1$ and $p_2$, $\psi(p_1)=\psi_1$ and $\psi(p_2)=\psi_2$, where the quasi-geostrophic vorticity equation for the mid-latitude $\beta$-plane (\ref{eq:mod1}) - (\ref{eq:mod2}) is applied, while the vertical velocity $\omega$ is specified at level $p_{1.5}$, where the thermodynamic energy equation (\ref{eq:mod3}) is valid. 

The model is described by the following equations in terms of the barotropic stream function $\psi_M=(\psi_1+\psi_2)/2$, baroclinic stream function $\psi_T=(\psi_1-\psi_2)/2$, and temperature $T$:
\begin{equation}
  \frac{\partial}{\partial t}(\nabla^2\psi_M)=-J(\psi_M,\nabla^2\psi_M+\beta y)-J(\psi_T,\nabla^2\psi_T)-r\nabla^2(\psi_M-\psi_T)+k_h\nabla^4\psi_M,
  \label{eq:mod1}
\end{equation} 

\begin{equation}
  \frac{\partial}{\partial t}(\nabla^2\psi_T)=-J(\psi_T,\nabla^2\psi_M+\beta y)-J(\psi_M,\nabla^2\psi_T)+r\nabla^2(\psi_M-\psi_T)+k_h\nabla^4\psi_T+\frac{f_0}{\Delta p}\omega,
  \label{eq:mod2}
\end{equation} 

\begin{equation}
  \frac{\partial}{\partial t}(T)=-J(\psi_M,T)+S_p\omega+r_R(T_e-T)+\kappa\nabla^2T
  \label{eq:mod3}
\end{equation} 
In the above, we expressed the advection in terms of the Jacobian operator defined as $J(A,B)=\frac{\partial A}{\partial x}\frac{\partial B}{\partial y}-\frac{\partial A}{\partial y}\frac{\partial B}{\partial x}$. $S_p$ represents the static stability parameter \citep{holton2004}. We define the stability parameter $S=\frac{R S_p \Delta p}{2 f_0^2}=L_D^2$, where $L_D$ is the Rossby radius of deformation. The name and values of model parameters are listed in Table \ref{tab:par}.

The vertical velocity is set to 0 at the top level, $\omega_{0}=0$, and is defined through Ekman pumping at the surface level, $\omega_{2.5}=\frac{\Delta p}{f_0} 2r \nabla^2 \psi_2$, which parameterises the dissipative processes occurring in the boundary layer. Subgrid-scale processes are represented by momentum and heat diffusion terms. The system is driven by a Newtonian cooling term that involves the restoration temperature field:
\begin{equation}
  T_e=\frac{\Delta T}{2}\cos\frac{\pi y}{L_y}.
  \label{eq:Te}
\end{equation} 
$\Delta T$ denotes the forced meridional temperature difference, and quantifies the external forcing in the model. In the performed simulations, no time-dependence of $\Delta T$ is assumed, with the aim of creating time series of a deterministic equivalent of a stationary process. If $\Delta T$ is sufficiently large, the system reaches a steady state featuring a turbulent atmospheric flow with sensitive dependence on initial conditions. The physical processes responsible for limited predictability are in general the baroclinic and barotropic instability. The Newtonian cooling provides the so-called baroclinic forcing to the system, and activates a set of energy exchanges and transformations summarised by the framework of the Lorenz energy cycle. See discussion in \citet{holton2004}. 

Using hydrostatic approximation \citep{holton2004}, we obtain for our vertical discretization $T=\frac{2f_0}{R}\psi_T$. Thus, the three model equations (\ref{eq:mod1}) - (\ref{eq:mod3}) can be reduced to two equations with two variables $\psi_M$ and $\psi_T$. 
The model equations can be transformed \citep{schubert2015} into a non-dimensional form using the scaling factors in Table \ref{tab:par}. \textcolor{black}{In the following, we use non-dimensional quantities, if not indicated otherwise.} As mentioned, the channel is periodic in the $x$-direction. At the meridional boundaries, we set the meridional velocity and the zonally-integrated zonal velocity to 0, $v=\frac{\partial \psi}{\partial x}=0$ and $\left. \int_0^{2\pi/a} \mathrm{d}x \frac{\partial \psi}{\partial y} \right|_{y=0,\pi}\!=0$. For these boundary conditions, the solution of the model equations is
\begin{equation}
  \psi(x,y,t)=\sum_{k,l=1}^{N_x,N_y}\left(\psi^r(k,l,t)\mathrm{cos}(akx)+\psi^i(k,l,t)\mathrm{sin}(akx)\right)\mathrm{sin}(ly)+\sum_{l=1}^{N_y}\psi^r(0,l,t)\mathrm{cos}(ly),
  \label{eq:sol}
\end{equation}
with $\psi$ denoting the non-dimensional stream function, index $r$ represents the real and $i$ the imaginary coefficients. We apply a spectral cut-off at $N_x=N_y=16$ in both $x$ and $y$ directions. Hence, the total dimension of the model phase space is $2N_y(2N_x+1)=1056$.
\subsection{Methods}

We substitute Eq.~(\ref{eq:sol}) into the evolution equations, perform a Galerkin projection, and eventually integrate numerically the non-dimensional model equations in spectral space using the fourth-order Runge-Kutta scheme. We perform simulations with two different forced meridional temperature differences, $\Delta T=133$ K and $\Delta T=40$ K. 
In the case of strong forcing ($\Delta T=133$ K), the system has a Kaplan-Yorke (KY) dimension of 585.95 with 222 positive Lyapunov exponents, so that $d_u=222$, $d_n=2$, and $d_s=361.95$. Note that the presence of a second neutral direction is related to the existence of a rotational symmetry in the system and to the fact that we consider a spectral model. This feature is of little relevance for the analysis below. We produce stationary time series of 96,576 years with a time step of 0.7 hours. In the case of weak forcing ($\Delta T=40$ K), the system has a KY dimension of 39.31 with 17 positive Lyapunov exponents, so that $d_u=17$, $d_n=2$, and $d_s=20.31$. We produce stationary time series of 485,760 years with a time step of 2.8 hours. The spectral coefficients $\psi^{r/i}$ of the stream functions are recorded every 5.5 hours with either forcings. The Lyapunov exponents are obtained by the same simulation code as the one used by \citet{schubert2015}, based on the method of \citet{benettinetal1980}.

\begin{table}[h]\small
  
  \caption{List of symbols and parameter values  for the QG model, Eq. (\ref{eq:mod1}) -- (\ref{eq:sol}).}
  \centering
  \resizebox{\textwidth}{!}{\begin{tabular}{l l l l l l}
    
    \toprule
    \textit{Variable}				& Symbol 	& Unit 									& 				& Scaling factor\\
    \midrule
    Stream function				& $\psi$	& $\mathrm{m}^2 \mathrm{s} ^{-1}$					& 				& $L^2 f_0$ \\
    Temperature					& $T$		& K									& 				& $2f_0^2L^2/R$ \\
    Velocity					& $\mathbf{v}$	& $\mathrm{m s}^{-1}$							& 				& $L f_0$ \\
    Energy					& $e$		& $\mathrm{J kg}^{-1}$							& 				& $L^2 f_0^2$ \\
    \midrule
    \textit{Parameter} 				&Symbol		& Dimensional Value							& Non-dimensional value 	& Scaling Factor \\
    \midrule
    Forced meridional temperature difference	& $\Delta T$	& 133 \& 40 K								& 0.188 \& 0.0564 		& $2f_0^2L^2/R$ \\
    Ekman friction 				& $r$ 		& $2.2016 \times 10^{-6}$ $\mathrm{s}^{-1}$ 				& 0.022 			& $f_0$ \\
    Eddy-momentum diffusivity 			& $k_h$ 	& $10^5$ $\mathrm{m}^2 \mathrm{s}^{-1}$					& $9.8696 \times 10^{-5}$ 	& $L^2 f_0$ \\ 
    Eddy-heat diffusivity 			& $\kappa$ 	& $10^5$  $\mathrm{m}^2 \mathrm{s}^{-1}$ 				& $9.8696 \times 10^{-5}$ 	& $L^2 f_0$ \\ 
    Thermal damping 				& $r_R$ 	& $1.157 \times 10^{-6}$ $\mathrm{s}^{-1}$ 				& 0.011				& $f_0$ \\
    Stability parameter				& $S$	 	& $3.33 \times 10^{11}$ \& $2.52 \times 10^{11}$ $\mathrm{m}^2$ 	& 0.0329 \& 0.0247 		& $L^2$ \\
    Coriolis parameter 				& $f_0$ 	& $10^{-4}$ $\mathrm{s}^{-1}$ 						& 1 				& $f_0$ \\
    Beta ($df/dt$)				& $\beta$ 	& $1.599 \times 10^{-11}$ $\mathrm{m}^{-1} \mathrm{s}^{-1}$ 		& 0.509 			& $f_0/L$ \\
    Aspect ratio 				& $a$ 		& 0.6896  								& 0.6896 			& - \\
    Meridional length 				& $L_y$ 	& $10^7$  m 								& $\pi$ 			& $L$ \\
    Zonal length 				& $L_x$ 	& $2.9 \times 10^7$  m 							& $2\pi/a$ 			& $L$ \\
    Specific gas constant			& $R$		& $ 287.06$ $\mathrm{J} \mathrm{kg}^{-1} \mathrm{K}^{-1}$		& 2 				& $R/2$ \\
    Vertical pressure difference		& $\Delta p$	& 500 hPa								& 1 				& $\Delta p$ \\
    Time scale					& $t$		& $10^4$ s								& 1 				& $1/f_0$ \\
    Length scale				& $L$		& $10^7/\pi$ m								& 1 				& $10^7/\pi$ \\
    \bottomrule
  \end{tabular}}
  \label{tab:par}
\end{table}

The spectral output of the model is transformed into the grid point space using Fast Fourier Transform resulting $n_x \times n_y$ grid points with $n_x=n_y=36$ in the $x$ and $y$ directions. We refer to the grid points by indices $(i_{x}, i_{y})$ where $i_{x}=i_{y}=0,...,35$. We analyse extremes of total energy observables defined in non-dimensional form below. For our extreme value analysis we consider only the ``mid-latitudes'' of the QG model, which we define as the region between the latitudes $i_{y}=9$ and $i_{y}=26$, i.e., the latitudinally central 0.5 fraction of the whole domain. The total energy is the sum of the kinetic energy of the lower and upper layers and of the available potential energy:
\begin{equation}
   e(i_x,i_y,i_t)=e_{k_1}(i_x,i_y,i_t)+e_{k_2}(i_x,i_y,i_t)+e_{p}(i_x,i_y,i_t),
   \label{eq:et}
\end{equation}
where $i_t$ represents the discrete time coordinate.
The components of the right side of (\ref{eq:et}) are defined for each grid point as:
\begin{equation}
   \ e_{k_1}(i_x,i_y,i_t)=\frac{1}{2}\left(u_1(i_x,i_y,i_t)^2+v_1(i_x,i_y,i_t)^2\right),
   \label{eq:ek1}
\end{equation}
\begin{equation}
   \ e_{k_2}(i_x,i_y,i_t)=\frac{1}{2}\left(u_2(i_x,i_y,i_t)^2+v_2(i_x,i_y,i_t)^2\right),
   \label{eq:ek2}
\end{equation}
\begin{equation}
   \ e_{p}(i_x,i_y,i_t)=2\lambda^2\psi_T(i_x,i_y,i_t)^2,
   \label{eq:ep}
\end{equation}
with the zonal component of the horizontal velocity $u_1=-\frac{\partial \psi_1}{\partial y}$, $u_2=-\frac{\partial \psi_2}{\partial y}$, the meridional component of the horizontal velocity $v_1=\frac{\partial \psi_1}{\partial x}$, $v_2=\frac{\partial \psi_2}{\partial x}$, and $\lambda^2=1/(2S)$.

We obtain the zonally-averaged energy by taking the zonal average of the local energy (Eq. (\ref{eq:et})):
\begin{equation}
  e_{\mathrm{z}}(i_y,i_t)=\frac{1}{n_x}\displaystyle\sum_{i_x=0}^{n_x-1} e(i_x,i_y,i_t),
  \label{eq:ez}
\end{equation}
and the average mid-latitude energy by averaging the local energy over the area corresponding to the mid-latitudes: 
\begin{equation}
  e_{\mathrm{ml}}(i_t)=\frac{2}{n_x n_y}\displaystyle\sum_{i_x=0}^{n_x-1} \displaystyle\sum_{i_y=9}^{26} e(i_x,i_y,i_t).
  \label{eq:eml}
\end{equation}
The energy observables are analysed in their non-dimensional form. The physical values expressed in units of J/Kg ($\mathrm{J/m}^2$) can be obtained by multiplying the non-dimensional values by the factor $L^2 f_0^2=1.013 \times 10^5$ ($L^2 f_0^2 \Delta p/g=5.164 \times 10^9$).

Although we record the model output, as stated above, every 5.5 hours, we save only the maximum values over one month in the case of strong forcing and over three months in the case of weak forcing. We estimate the GEV and GP parameters based on block maxima and threshold exceedances obtained from the monthly, respectively  3-monthly, maxima series. Such an operation has no effect on the subsequent GEV analysis. Instead, it might modestly impact the GP analysis, as some above threshold events might be lost, because they could be masked by a larger event occurring within the same 1-month or 3-months period. Nonetheless, since we consider very high thresholds and an extremely low fraction of events, the risk of losing information is negligible. The GEV and GP parameters are inferred by maximum likelihood estimation (MLE), as described by \citet{coles2001}. We estimate the GEV and GP parameters, as well as the confidence intervals, using the MATLAB functions \texttt{gevfit} and \texttt{gpfit}. The computed 
confidence intervals contain the true value of the parameters with a probability of $95\%$. The auto-correlation coefficients and histograms are obtained based on 1000 years of the ``raw'' simulated time series.

\section{Extreme value statistics in the QG model}

As mentioned before, the simulations are performed using two different configurations, where the value of the parameter $\Delta T$, describing the baroclinic forcing, is set to 133 K and 40 K, respectively. As we see below, in the case of strong forcing we find good agreement between the results of the statistical inference and the theory, for local observables at least, even if the speed of convergence of the estimated shape parameters (not predicted by the theory), to the value that is predicted by the theory is rather diverse among the considered observables. In the case of weak forcing and the resulting weakly turbulent behaviour, the results of the statistical inference analysis are in not so good agreement with the theory, and we find that for the different observables the shape parameter estimates have various non-monotonic dependence on the block size. We will investigate possible reasons for such a behaviour.  

\subsection{Strong forcing ($\Delta T=133$ K)}
\label{sec:tg20}

Before presenting the results related to the statistics of extreme events, we outline some general statistical properties of the analysed observables. As emphasised in Sec.~1 and 2, correlations have an effect on the convergence of the distribution of block maxima (threshold exceedances) to the GEV (GP) distribution. The auto-correlation coefficient for a stationary signal $f(t)$ can be defined as follows:
\begin{equation}
 \rho_f(l)=\frac{\mathrm{E}[(f(t)-\mu_f)(f(t+l)-\mu_ f)]}{\sigma_f^2},
 \label{eq:ac2}
\end{equation}
where $t$ represents the time and $l$ the time lag, $\mu_f$ denotes the mean and $\sigma_f$ the variance of $f(t)$ \citep{dunn2010}.

By taking the ergodic hypothesis, we estimate the auto-correlation coefficient for the local energy according to (\ref{eq:ac2}) and obtain for each grid point $\rho_e=\rho_e(i_x,i_y,l)$. We calculate the integrated auto-correlation time scale \citep{franzkeetal2005} $\tau_e=\tau_e(i_x,i_y)$ according to $\tau_e=\sum_{l=0}^{n_l}|\rho_e|$ \footnote{If the decay of the auto-correlation is exponential, the integrated auto-correlation time scale is equal to the e-folding time.}. We set $n_l=604$ (corresponding to about 140 days) as an upper limit for the integration in order to avoid the noisy tail of the auto-correlation coefficient. We proceed the same way in the case of the zonally-averaged and average mid-latitude energy to obtain $\tau_{e_z}(i_y)=\sum_{l=0}^{n_l}|\rho_{e_z}(i_y,l)|$ and  $\tau_{e_{ml}}=\sum_{l=0}^{n_l}|\rho_{e_{ml}}(l)|$. The integrated auto-correlation time scales, expressed in days, are shown in Fig.~\ref{fig:stat}~(a). As expected, the weakest auto-correlations are recorded for the local observables, yielding about 1-2 days. Because of the information propagation along parallels due to the prevailing zonal winds, the zonal average time series are impacted by a low-pass filtering as result of averaging along a latitudinal band, thus the correlations become stronger. For these zonally-averaged observables, as opposed to the local ones, the integrated auto-correlation changes substantially with latitudes. We observe a minimum in the middle of the channel ($\approx 3.5$ days) and an increase outwards to the boundaries ($\approx 15$ days). Through averaging over the area of mid-latitudes, the zonally-averaged time series with different properties are merged together. The resulting time series has an integrated auto-correlation time scale of about $3.6$ days. Note that if in a time series of length $N$ with reasonably fast (e.g. similar to exponential) decay of correlations the integrated auto-correlation time scale is $\tau$, then one can deduce that the time series has approximately $N \times \Delta t/\tau$ effectively independent entries, where $\Delta t$ is the time interval. This can have important effects in determining when the asymptotic behaviour of the EVT statistics is valid.

Fig. \ref{fig:stat} (b) -- (d) illustrates the  histograms and the approximated probability density functions (PDFs) of our observables. Although, due to the $\beta$-effect, the dynamical properties of the flow as a function of latitudes are not exactly symmetric with respect to the meridional middle of the channel, our estimations of statistical quantities and density functions exhibit approximate meridional symmetries. Therefore, in the case of the local and zonally-averaged observables only half of the channel's meridional extension (at every second latitude) is shown. The strongest skewness and the longest right tails are observed in the case of the PDFs of the local observables. After spatial averaging, the PDFs become more symmetric and almost similar to a Gaussian distribution (which would look like a parabola on a semi-logarithmic scale), according to the central limit theorem.

\begin{figure}[ht!]
  \centering
  \includegraphics[width=0.8\textwidth]{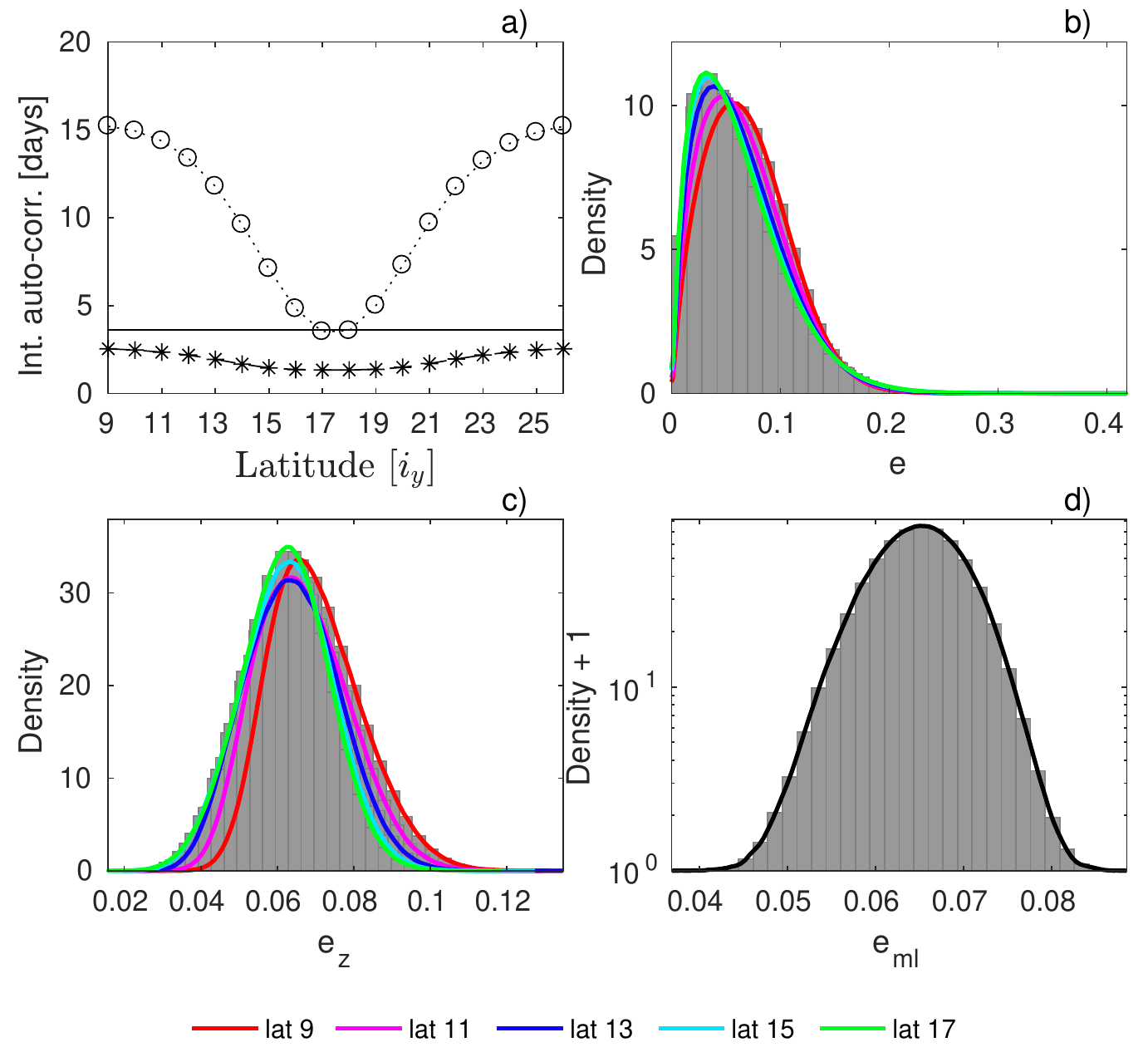}
  \caption{Statistical properties of the total energy for $\Delta T=133$ K. (a) Integrated auto-correlation time scales: zonal average of $\tau_e$ (dashed line with star markers), $\tau_{e_z}$ (dotted line with circle markers), $\tau_{e_{ml}}$ (continuous line); histograms of the (b) local, (c) zonally-averaged, and (d) average mid-latitude observables. In the case of (b) -- (d), the continuous lines show the approximation of the PDFs by kernel smoothing (\textit{ksdensity} function in MATLAB), the colours mark different latitudes according to the legend.} 
  \label{fig:stat}
\end{figure}

In what follows, we present the results of the extreme events analysis starting with the local observables. We first discuss the convergence of the shape parameter for GEV and GP, then the convergence of the GP modified scale parameter (to be introduced below), and, at the end, the convergence of return levels. Taking advantage of the fact that statistics are uniform in the zonal direction, we concatenate the monthly maxima series for every second longitude one after the other in the $x$-direction, thus increasing the data length to about $1.7 \times 10^6$ (from about $9.6 \times 10^4$) years. Therefore, we can estimate the GEV and GP shape parameters for larger block sizes and higher thresholds than in the case of the zonally-averaged or average mid-latitude observables. Although the time series at every second longitude are correlated with each other, the correlation almost vanishes at the block size of 8 years, being below 0.15 at every latitude. In other words, correlations are very weak at extreme levels, which is according to \citet{coles2001} the only important condition for the GEV limit laws to apply in the case of a stationary process. Block sizes smaller than 8 years are not relevant for our analysis, since (as presented below) much larger ones are needed to approach the theoretical shape parameter. In the case of the POT approach, we use the same argument of choosing very high thresholds, above which the correlations are extremely weak.

\textcolor{black}{The theory discussed in Sec.~\ref{sec:theory} indicates that the true (asymptotic) GEV shape parameter is given by $\xi_\delta$, as expressed by Eq.~(\ref{eq:xi}), which corresponds to approximately -0.002, and is indicated by the straight line in Fig.~\ref{fig:shape20}. Note that the range of the theoretical values derived  taking into consideration possible geometrical degeneracies, according to what is described in Sec.~\ref{sec:theory}, is too small to be visible in this case. We define the \textit{precision} $P(n)$ of estimation by considering half of the width of the 95~\% maximum likelihood confidence interval.  It is more common to define the precision by  the standard deviation of the estimates. For a Gaussian distribution the 95 \% confidence interval is larger than the standard deviation by a factor of approximately $2$. However, this distinction does not matter for our purposes. Besides, we have a \textit{single} estimate only, and so we can obtain only the confidence interval not the standard deviation of the distribution of estimates.
 Additionally to the precision, we define the \textit{trueness} of a \textit{single} estimate by the distance between $\xi_{\delta}$ and $\xi(n)$: $T(n)=|\xi_{\delta}-\xi(n)|$. Note that the latter is different from the usual definition in that the reference from which we measure the distance is $\xi_{\delta}$, not the true value of the distribution from which the BM data is drawn. In fact, strictly, the BM data is not drawn from a GEV distribution that we are fitting, and hence we cannot even talk about the true value of an underlying GEV distribution. We emphasise that our interest is the convergence to the asymptotic value, which is why we take a reference value in our definition other than customary. Accordingly, we shall refer to the `bias' of the estimator, again different from customary, as the expected trueness. We remark that, since our estimates are obtained based on one realisation instead of several realisation yielding a distribution of estimates, our trueness $T$ approximates the bias of the estimates as long as $P\ll T$. We emphasise that we are able to calculate $T$ here because we know the true $\xi_{\delta}$; this is not the case in practice when facing just a measured time series. Obviously, we aim at obtaining a joint optimisation by having a bias and a precision as small as possible. Clearly, optimality requires a compromise between these two requirements. When we apply the BM method and increase the block size $n$, the number of blocks and of BM decreases, thus the estimation of $\xi(n)$ becomes more and more uncertain, and P increases monotonically. At the same time, for increasing $n$ we expect a (not necessarily monotonic) convergence of our estimated shape parameter to the true value, so that the actual bias should (on the long run) decrease with $n$. Clearly, instead, our approximation $T$ decreases only until a certain block size, above which it becomes more uncertain with increasing values of $n$, because less BM are available. We choose as optimal block size $n=n^*$ the smallest block size for which the estimate of the bias is lower than the estimate of the precision $n^*=\min(n;T(n)<P(n))$. On the scale of variation ranges of $P(n)$ and $T(n)$ we have $T(n^*)\approx P(n^*)$. With this we obtain a single number that can quantify the \textit{accuracy} of estimation. This measure of accuracy provides here a basis for comparing different observables with regard to the speed of convergence, or a basis for assessing the degree of non-uniformity of estimates of various observables of interest (in terms of the range of accuracy values), as a finite-data-size deviation form the uniformity predicted by theory. We have verified that the optimal choice for $n^*$ is virtually unchanged when we use an alternative definition of the accuracy such as $T^2+P^2$, borrowing an idea concerning the optimality of MLE estimators (not shown).}

First we assess the uniformity for the local observables. Figure~\ref{fig:shape20}~(a) shows the GEV shape parameter estimates against exponentially increasing block sizes of $n=2^{i}$ years ($i=-2,-1,...,13$), for different latitudes. The estimated GEV shape parameters $\xi(n)$ seem to converge monotonically for every latitude to $\xi_{\delta}$. The monotonic convergence is pointed out also in panel (b) in terms of $T(n)$. In this diagram we display $P(n)$ too, by which we can determine the optimal $n^*$ and the accuracies of estimation. These accuracies, depending on the latitude, have a range of $5\times10^{-3} - 2\times10^{-2}$. At the same time, the value of $n^*$ ranges from several tens of years to a few hundreds of years depending on the latitude we are considering. This is unsurprising, because the speed of convergence to the asymptotic level is not universal. As a consequence, when finite block sizes are considered, extremes of different observables can feature rather distinct properties. The slow convergence suggests that customary choices like yearly maxima are not always good enough for an accurate modelling of extremes. Figure~\ref{fig:shape20} (c), giving a different view of the same data seen in panel (a), illustrates the estimated GEV shape parameter as a function of latitudes for various block sizes. For small block sizes, we observe a slight latitudinal dependence of the shape parameter. This latitudinal structure flattens as one increases the block size, and the estimated shape parameters get closer to the theoretical value. According to Fig.~\ref{fig:shape20}~(c), universality emerges as we approach the asymptotic level.

To assess the goodness of fit, we perform a one-sample Kolmogorov-Smirnov-test (KS-test) \citep{massey1951} at 5~\% significance level using the MATLAB function \texttt{kstest}. \textcolor{black}{We remark that the KS-test is performed in case of each block size based on the whole BM data, meaning that this amount of data decreases as we increase the block size}. The shape parameter values for which the KS-test p-value $p$ is above 0.05 (i.e., the hypothesis that the distribution of BM is a GEV distribution cannot be rejected) are marked by circle markers in Fig.~\ref{fig:shape20}~(a). We define $n_{KS}$ as the smallest block size for which $p>0.05$, $n_{KS}=\min(n;p>0.05)$, and $\xi_{KS}=\xi(n_{KS})$. Figure \ref{fig:shape20} points out that the KS-test suggests a good fit already at smaller block sizes than the optimal block size, $n_{KS}<n^*$, and for lower shape parameter values than the best estimate, $\xi_{KS}<\xi(n^*)$. Thus, a very important conclusion is that the p-value of the KS-test is not an appropriate measure for the convergence to the limiting distribution. More precisely, it indicates that we have indeed agreement with a member of the GEV family of distributions, but we cannot say what is the error from the asymptotic value of the parameters. We emphasise that $n_{KS}$, just like $n^*$, depends on the time series length, and it would be even smaller if shorter time series were considered. This implies that in the case of applications with less data, the results of the KS-test are even less reliable. The misleading property of p-values was also shown by \citet{bodai2017}, who studied the convergence to the GEV distribution of extremes of site variables in the Lorenz 96 model, and found p-values above the significance level in cases where the theoretical prediction did not even apply, and the shape parameter did not converge. The goodness-of-fit test was in the mentioned study a Pearson's chi-squared test. Misleading p-values based on the KS-test were pointed out also by \citet{faranda2011} in the case of the BM approach in simple systems. \textcolor{black}{A slow convergence of the estimated GEV shape parameters and a poor quality of diagnostic tools (return level and quantile plots) in case of small block sizes were also found by \citet{vannitsem2007} in case of local temperature extremes in a three-layer QG model with orography}.

Figure~\ref{fig:shape20} (d) illustrates the GP shape parameter estimates as a function of decreasing exceedance ratio (the fraction of above-threshold data) $r$, which is equivalent with an increasing threshold. To ensure direct comparability between the BM and POT approaches of EVT, sample values of the threshold are chosen corresponding to the sample values of the block size, in such a way that $r=\frac{1}{n\times m_y}$, where $m_y$ is the data amount in a year. Thus the number of threshold exceedances is equal to the number of block maxima. By comparing the GP shape parameter (Fig.~\ref{fig:shape20}, d~--~f) with the GEV shape parameter (Fig.~\ref{fig:shape20}, a~--~c), we generally observe the same characteristics. More precisely, the changes of the GP shape parameter as a function of exponentially decreasing exceedance ratio are very similar to the variation of the GEV shape parameter according to an exponentially increasing block size. Both GEV and GP shape parameters seem to converge to $\xi_\delta$. This is also consistent with theoretical results according to which the two distributions are asymptotically equivalent \citep{coles2001,lucarini2014}. However, we expect that in case of finite block sizes (i.e., in case of every practical application) differences might emerge in the estimates of the GEV and GP shape parameters. Although, in the case of consistent estimations one would expect that at large block sizes, corresponding to low exceedance ratios, the difference between them should be small, as it is the case for our estimators. Besides the mentioned similarities, we observe some differences between the estimates of the GEV and GP shape parameters. These differences concern, for example, their latitudinal dependence (less pronounced in case of the GP shape parameter) or the width of the confidence intervals (larger in case of the GP shape parameter, indicating larger estimation uncertainty). The most relevant difference is, however, that the GP shape parameter seems to converge faster to $\xi_\delta$. This is unsurprising as in many applications it is usually suggested to use the POT over the BM method as the former is less data-hungry and provides (usually) a faster convergence \citep{lucarini2016}.

\begin{figure}[ht!]
  \centering
  \includegraphics[width=1\textwidth]{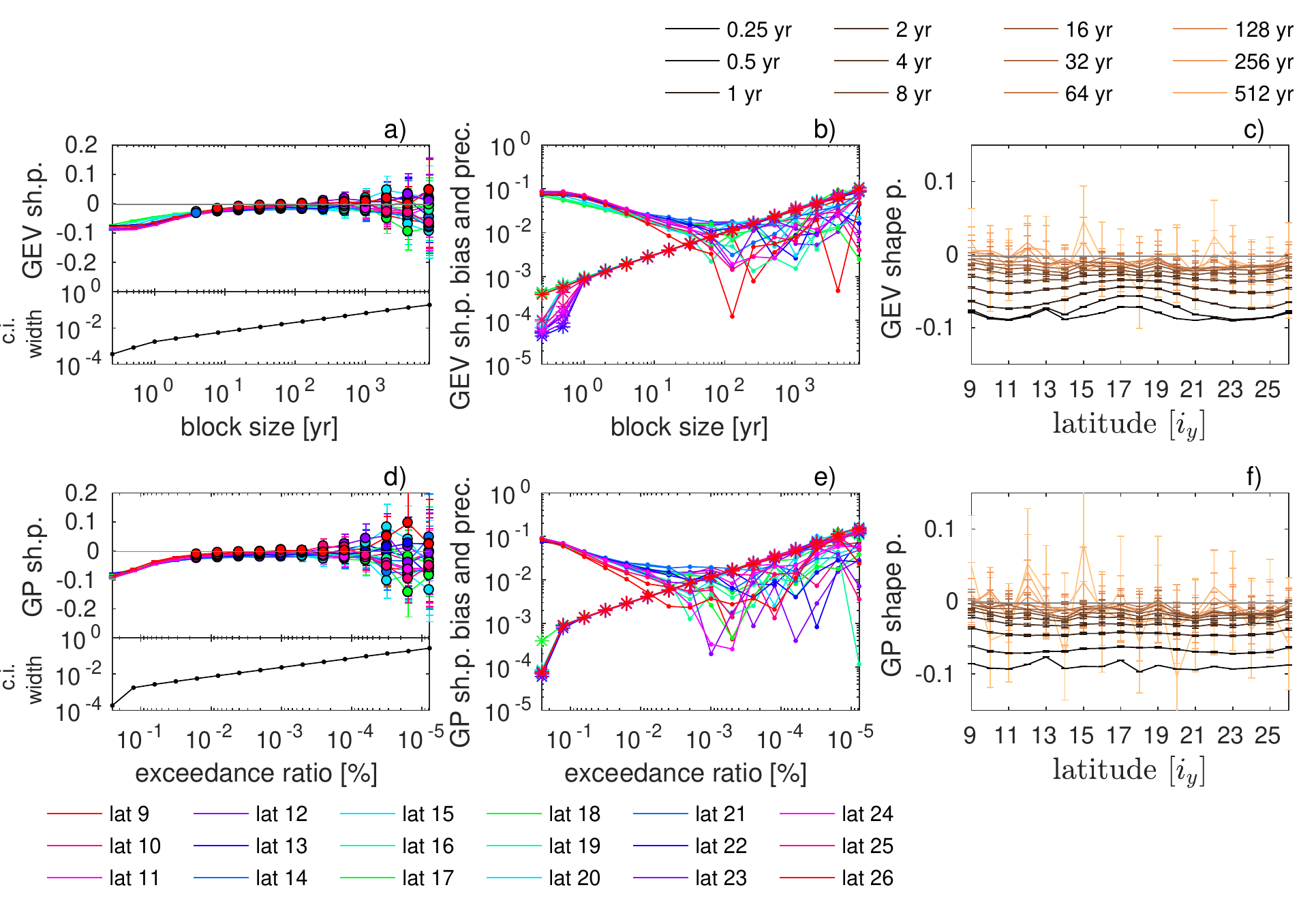}
  \caption{GEV and GP shape parameters as well as bias and precision estimates in case of the local observables, for $\Delta T=133$ K. (a) GEV and (d) GP shape parameter estimates as functions of the block size and exceedance ratio. The circle markers point out shape parameter values, for which the p-value of the KS-test is above 0.05. Lower $y$-axis: zonal mean of the maximum likelihood 95~\% confidence interval widths. (b) and (e): Estimates for the bias (dot markers) and precision (star markers) of the shape parameter. (c) GEV and (f) GP shape parameter estimates as functions of the latitude. The grey, horizontal line illustrates the theoretical shape parameter of -0.002. The range of theoretical values resulting from taking into consideration possible geometrical degeneracies is invisible in this case. The error bars show the 95~\% confidence intervals of the MLE. Different colours represent different latitudes (a, b, d, e) or different block sizes (c) or exceedance ratios (f).} 
  \label{fig:shape20}
\end{figure}

We perform another test to check whether the GP distribution is a good approximation for the distribution of threshold exceedances based on our data, and consider the GP modified scale parameter. The GP scale parameter depends on the chosen threshold according to $\sigma_u=\sigma_{u_0}+\xi(u-u_0)$ \citep{coles2001}, where $\xi=\xi_{\delta}$ represents the asymptotic shape parameter, $u_0$ is the lowest threshold at which the GP distribution is a reasonable model for exceedances, and $u$ represents any other threshold $u>u_0$. The scale parameter can be reparameterized yielding the modified scale $\hat\sigma=\sigma_{u_0}-\xi u_0=\sigma_u-\xi u$, which should converge to a nonzero value. Figure \ref{fig:sclo20} illustrates the modified scale parameter estimate (calculated based on the finite-size GP parameter estimates, i.e., taking threshold dependent GP shape parameter estimates instead of $\xi_\delta$) as a function of the exceedance ratio $r$. We observe estimates of $\hat\sigma$ relatively stable to further decreases of $r$ (for $r<r^*, r^*=\max(r; T(r)<P(r))$), which supports the conclusions drawn before that we are indeed close to asymptotic levels as required by EVT. Note that in this case there is no universality in the value of the modified scale parameter, as for stochastic variables one has that the upper right endpoint of the distribution is given by $A_{max}=-\sigma_{u_0}/\xi+u_0=-\hat\sigma/\xi$. Such an endpoint is clearly observable-specific.

\begin{figure}[ht!]
  \centering
  \includegraphics[width=0.6\textwidth]{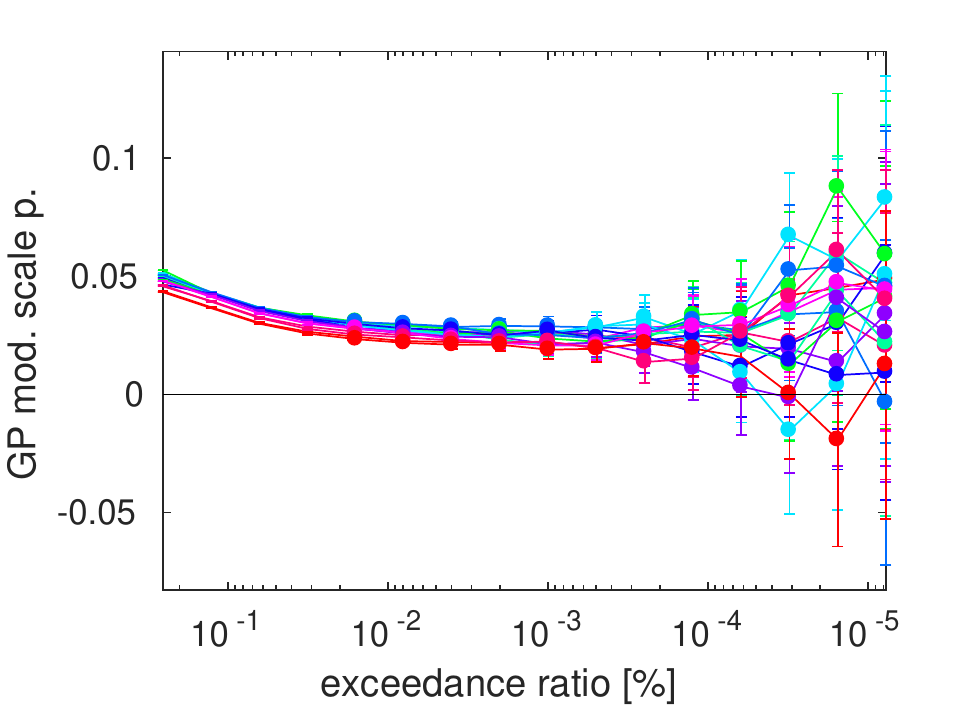}
  \caption{GP modified scale parameter estimates in case of the local observables, for $\Delta T=133$ K. The error bars show the 95~\% confidence intervals of the maximum likelihood estimation. Different colours represent different latitudes according to Fig. \ref{fig:shape20}.}. 
  \label{fig:sclo20}
\end{figure}

Having practical applications in mind, the BM and POT methods aim at obtaining statistical estimates of either return levels or expected return periods, for even unobserved extreme events. Figure~\ref{fig:rl20} (a) and (b) show GEV and GP return level plots for the local observables based on a fixed block size, $n=128$ years, and corresponding (as explained) \textcolor{black}{$r=5 \times 10^{-4}$~\%}, respectively, at five different latitudes (every second latitude from the southern meridional boundary to the channel centre). We compute the GEV return levels according to Eq.~(\ref{eq:rlgev}), the GP return levels based on Eq.~(\ref{eq:rlgpd}), and estimate the 95~\% confidence intervals using the delta method described by \citet{coles2001}. The GEV and GP return level plots look very similar, except two minor differences. One emerges simply from the different equations for the GEV and GP distributions, leading to slightly different definitions of return levels (as described in Sec. \ref{sec:theory} and in more details in \citealp{coles2001}) and affects short return periods; and the other one comes 
from the larger uncertainty in the estimation of the GP parameters compared to the GEV parameters, and results slightly wider confidence intervals in the case of the GP return levels. The main message of Fig. \ref{fig:rl20} (a) -- (b) is, however, that the GEV and GP return level estimates using the chosen $n$ and $r$ fit the empirical data quite well, which is in agreement with the results of the KS-test reported above. The 95~\% confidence intervals of the estimated return levels (continuous lines) contain the empirical return levels (dot markers) or are very near to them, except a few very high extremes at some latitudes. The return level is almost linear to the logarithm of the return period, showing the effect of a shape parameter very close to 0 (see Eq. (\ref{eq:rlgev}) and (\ref{eq:rlgpd})). 

If the GEV distribution is an adequate model for extreme events for a certain block size, one expects return levels with a certain return period not to change much any more with increasing block size. Figure \ref{fig:rl20} (c) -- (e) shows indeed that, above a certain block size, the estimated return levels for three different return periods ($10^3$, $10^4$, and $10^5$ years) are stable against further increase of $n$. But it also shows that the longer the return period, the slower the convergence. While in case of the $10^3$-years return period we obtain stable return level estimates already at $n_{KS}$, in case of $10^5$-years the return level estimates are still increasing for $n>n_{KS}$. Here we experience the practical effect of the issue mentioned above, namely, that the KS-test suggests a good fit even for $\xi_{KS}<\xi(n^*)$. This implies that the estimation of return levels with long return periods can be erroneous even if the KS-test does not reject the GEV distribution. We also notice that the return levels are underestimated if the block size is too small, and this underestimation is more severe in the case of return levels with longer return periods. We come to the same conclusion by considering the convergence of the GP return levels (not shown), as suggested already by the similarity between panels (a) and (b).

\begin{figure}[ht!]
  \centering
  \includegraphics[width=1\textwidth]{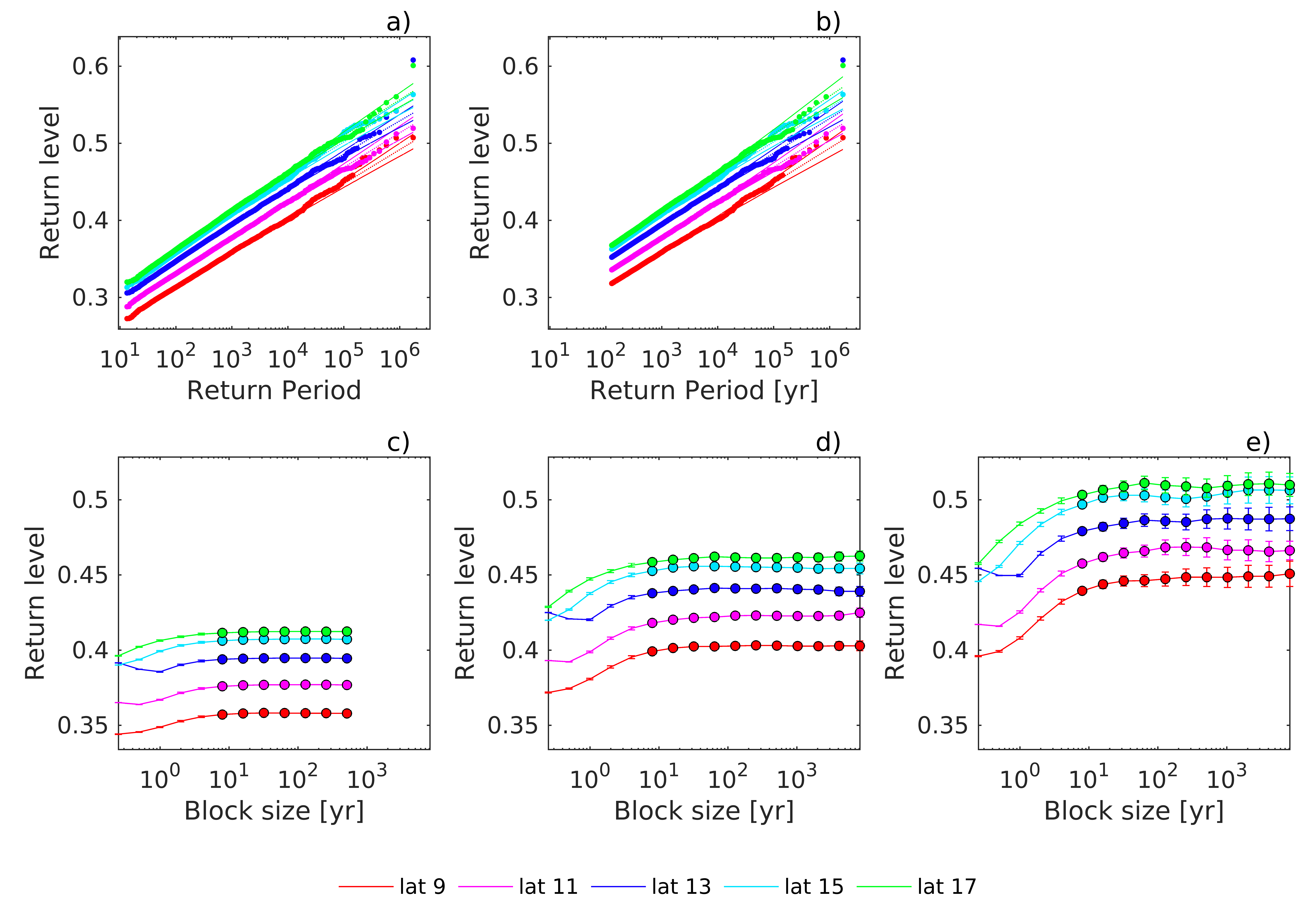}
  \caption{Return levels for $\Delta T=133$ K. (a) Return levels vs. return periods based on GEV parameters using a block size of 128 years and (b) based on GP parameters using an exceedance ratio of $5\times 10^{-4}$~\%. (Dotted lines: estimated return levels; continuous lines: 95~\% maximum likelihood confidence interval limits of the return level estimates; dot markers: empirical return levels.) GEV Return levels for (c) $10^3$-years, (d) $10^4$-years, and (f)  $10^5$-years return periods as functions of the block size. The error bars show the 95~\% confidence intervals of the MLE. The circle markers point out estimates for which the p-value of the KS-test is above 0.05. The colours mark different latitudes according to the legend.} 
  \label{fig:rl20}
\end{figure}

After having discussed in detail the convergence in case of the local observables, we proceed with the results for the zonally-averaged observables. Figure \ref{fig:shape20z} illustrates the GEV and GP shape parameters for the zonally-averaged observables (a, c, d, f) as well as the estimated bias and precision of the inferred shape parameters (b, e). As mentioned above, in case of the zonally-averaged observables we have shorter time series ($9.6 \times 10^4$ instead of $1.7 \times 10^6$ years). Because of this, results for the accuracies of estimates cannot be `fairly' compared to the accuracies found for local observables. Nevertheless, we produce the same type of diagrams suitable to determine the accuracies and show it in Fig. \ref{fig:shape20z} (b) and (e). Clearly, the range of accuracy values depending on the latitude and the maximal value of the accuracies (i.e, of the bias at the optimal block size) are both considerably larger than those for the local observables. What is fair to compare, however, is the range of biases for a certain block size where the confidence of the estimates is high, $P\ll T$, and the amount of data does not affect significantly the parameter estimate. In this regard, the zonal observables display a much larger non-uniformity regarding the shape parameter estimates. Otherwise, the estimates feature typically a monotonic change towards the theoretical value (up to at least the optimal block size), what can be seen as convergence.

Our observation that the estimated shape parameters depend strongly on the considered latitude has to do with the effect of serial correlation on the convergence to the limiting distribution. We obtain weak auto-correlations, fast convergence to $\xi_\delta$, and low bias in the middle of the channel, versus strong auto-correlations, slow convergence, and large bias at the margins of the channel. As already mentioned before, the stronger the serial correlation the less the number of uncorrelated data in a block, and the larger block sizes are needed in order to approach asymptotic levels (see also \citep{coles2001}). Thus the latitudinal structure of the GEV shape parameter estimates (Fig.~\ref{fig:shape20z},~c) is related to the one of the integrated auto-correlation time scale (Fig.~\ref{fig:stat},~a, dotted line with circle markers). By increasing the block size, this latitudinal structure flattens, the estimated shape parameters seem to approach $\xi_\delta$, and the confidence intervals contain $\xi_\delta$ in the case of several latitudes, especially the central ones. Nonetheless we note that, due to the presence of (relatively) large statistical uncertainty on the shape parameter, we cannot make more precise statements on the success of the analysis.

\begin{figure}[ht!]
  \centering
  \includegraphics[width=1\textwidth]{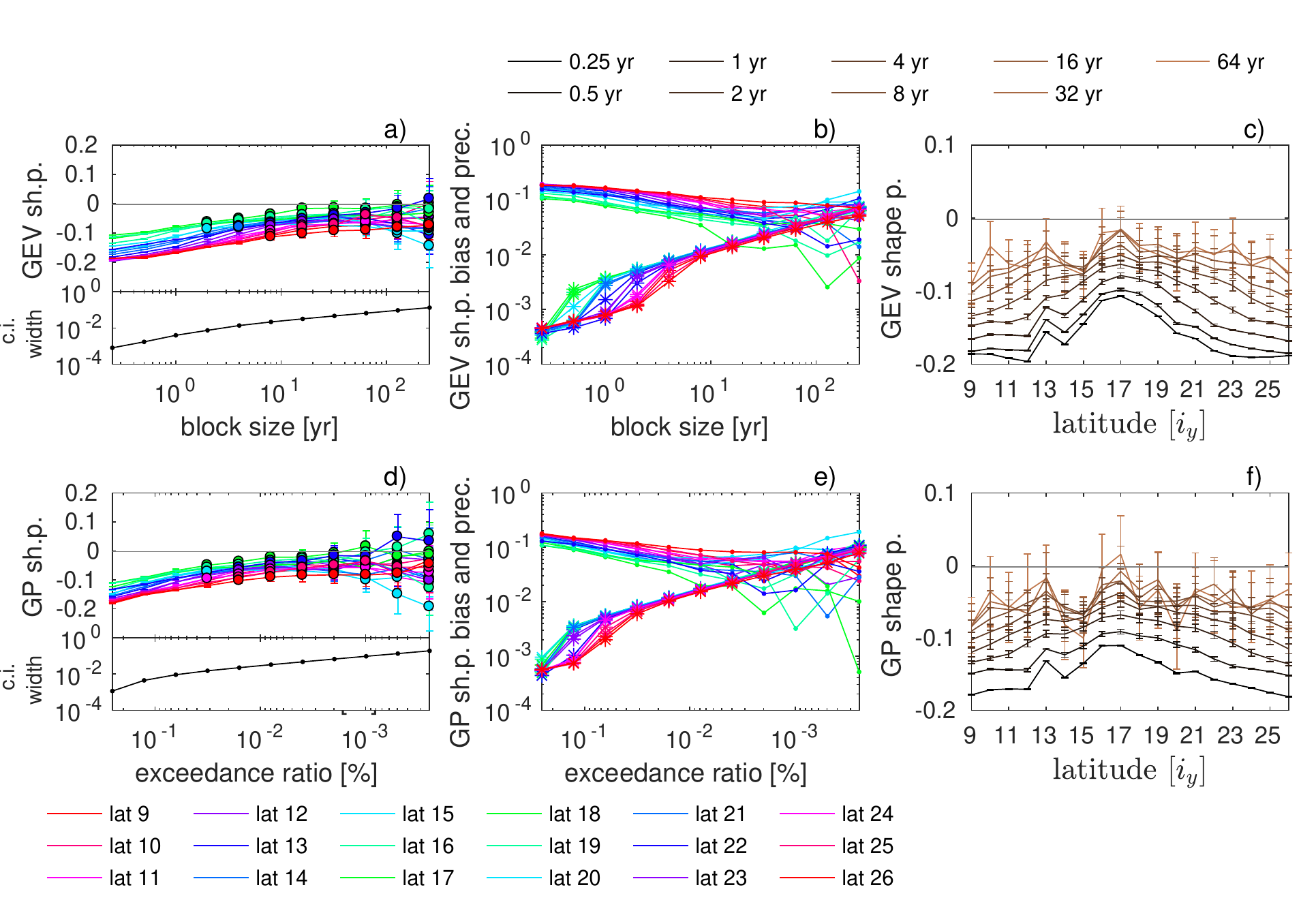}
  \caption{Same as Fig. \ref{fig:shape20}, but for the zonally-averaged observables.} 
  \label{fig:shape20z}
\end{figure}

We present now the analysis of extremes of the average mid-latitude observable. Figure \ref{fig:shape20ml} shows the GEV and GP shape parameter estimates for the average mid-latitude observable and their estimated bias and precision as a function of the block size and exceedance ratio, respectively. In case of the average mid-latitude energy, we have the same amount of data as in case of the zonally-averaged energy. Similarly to the zonally-averaged observables, the estimated GEV and GP shape parameters seem to approach the theoretical shape parameter, but, when more stringent definitions for selecting the extremes are used, the bias is relatively large, being about $4\times 10^{-2}$ at the optimal block size in the case of the GEV and about $7\times 10^{-2}$ at the optimal exceedance ratio in the case of the GP shape parameter. Again, also in this case, our analysis is limited by the amount of available data.

\begin{figure}[ht!]
  \centering
  \includegraphics[width=0.8\textwidth]{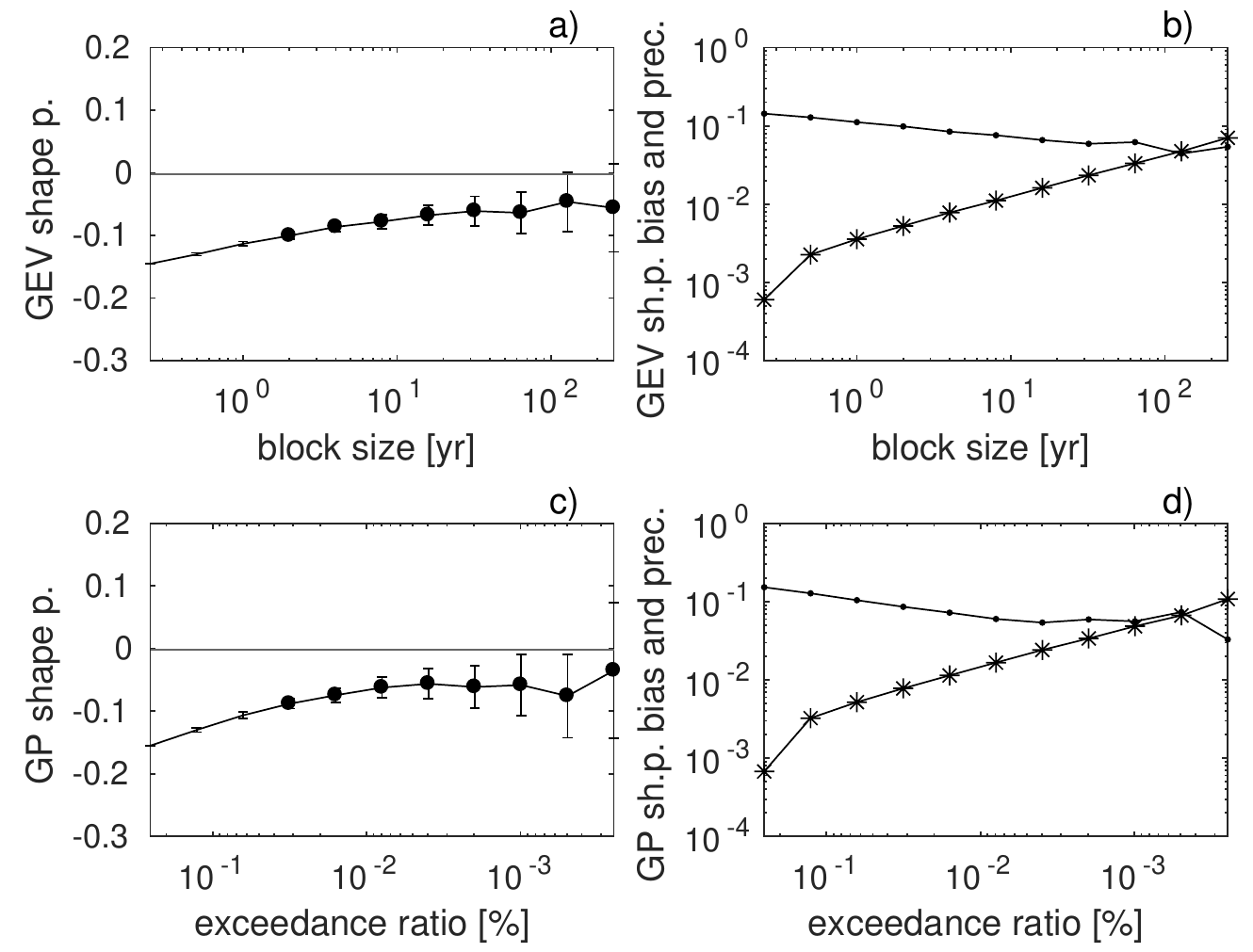}
  \caption{GEV and GP shape parameter as well as bias and precision estimates in case of the average mid-latitude observable, for $\Delta T=133$ K. (a) GEV and (c) GP shape parameters as functions of block size and exceedance ratio. The circle markers point out shape parameter values for which the p-value of the KS-test is above 0.05. The grey, horizontal line illustrates the theoretical shape parameter of -0.002. The range of theoretical values resulting from taking into consideration possible geometrical degeneracies is invisible in this case. The error bars show the 95~\% confidence intervals of the MLE. (b) and (d): Estimates of the bias (dot markers) and precision (star markers) of the shape parameter.} 
  \label{fig:shape20ml}
\end{figure}

In short, our numerical results do allow for conclusions regarding the universality of extremes, as predicted by the theory presented in Sec. \ref{sec:theory}. However, considering the most various observables one would typically see a non-uniformity in the finite-size shape parameter estimates simply because of their distinct convergence properties (not predicted by the theory). The observables that we found in our study to have the fastest converging shape parameter estimates are the the local observables at every latitude and the zonally-averaged observables at central latitudes, where the auto-correlation has a minimum. However, convergence is very slow, and is additionally slowed down by the presence of serial correlations in the time series. Thus, the estimated shape parameters are relatively far from the theoretical value in case of several latitudes of the zonally-averaged observables (especially marginal latitude exhibiting strong auto-correlations) and in case of the average mid-latitude observable. This slow convergence in combination with the finite size of the data makes the actual observation of the theoretical limit extremely difficult.

\subsection{Weak forcing ($\Delta T=40$ K)}

Before analysing the extreme events for weak forcing, we discuss some statistical (and dynamical) properties of our observables, which influence directly the statistics of extremes. Figure \ref{fig:stat6} (a) shows the integrated auto-correlation time scales for the three observables: local, zonally-averaged, and average mid-latitude energy. We compute the integrated auto-correlation time scale according to the method described in Sec.~\ref{sec:tg20} for the strong forcing. In the case of weak forcing, however, we set the time lag $n_l=1728$ (corresponding to about 400 days) as upper limit for the integration, according to the slow decay of the auto-correlation (especially in case of the zonally-averaged and average mid-latitude observables). The integrated auto-correlation time scales are substantially higher than for strong forcing: around 10 days in case of the local, about 30~--~48 days in case of the zonally-averaged observables, and approximately 45 days for the average mid-latitude observable. Figure \ref{fig:stat6} (b) shows the time series of the local observables at the central latitude $i_y=17$ (at two different longitudes $i_x=4$ and $i_x=19$) and suggests two alternating states of our system: one with strong fluctuations and another one with reduced fluctuations. Thus, it seems that our system exhibits a regime behaviour, which definitely supports the presence of strong correlations.

In contrast to the case of strong forcing, the zonal averages of the local energy observables show remarkable deviations from a Gaussian behaviour, even more than the PDFs of the local energy observables (Fig.~\ref{fig:stat6},~c~--~e). One has that the PDFs of the zonally-averaged observables typically have a marked skewness and very strong large kurtosis, and often contain rather pronounced ``shoulders", where smoothness is basically lost. The presence of large kurtosis indicates that there is significant positive spatial correlation of the energy along a longitude. The presence of skewness indicates that there is asymmetry between the occurrence of anomalies of either sign. Another particular property of the spatial energy field for weak forcing is the strong anti-correlation (especially in case of the zonally-averaged observables) between time series at central and marginal latitudes (not shown). Accordingly, the ``shoulders'' appear in different parts of the PDFs at different latitudes: on the left in the case of central latitudes and on the right in the case of marginal latitudes. We conclude that the regime behaviour is connected to non-trivial spatial structures, with the system living in a transitional range where one can still distinguish long-lived unstable waves amidst chaos. We note that such conditions are different from what is foreseen by the chaotic hypothesis, and, therefore, the statistics of extremes might not converge (according to our finite-sized data set) to what is predicted by the theory developed for Axiom A systems.

\begin{figure}[ht!]
  \centering
  \includegraphics[width=1\textwidth]{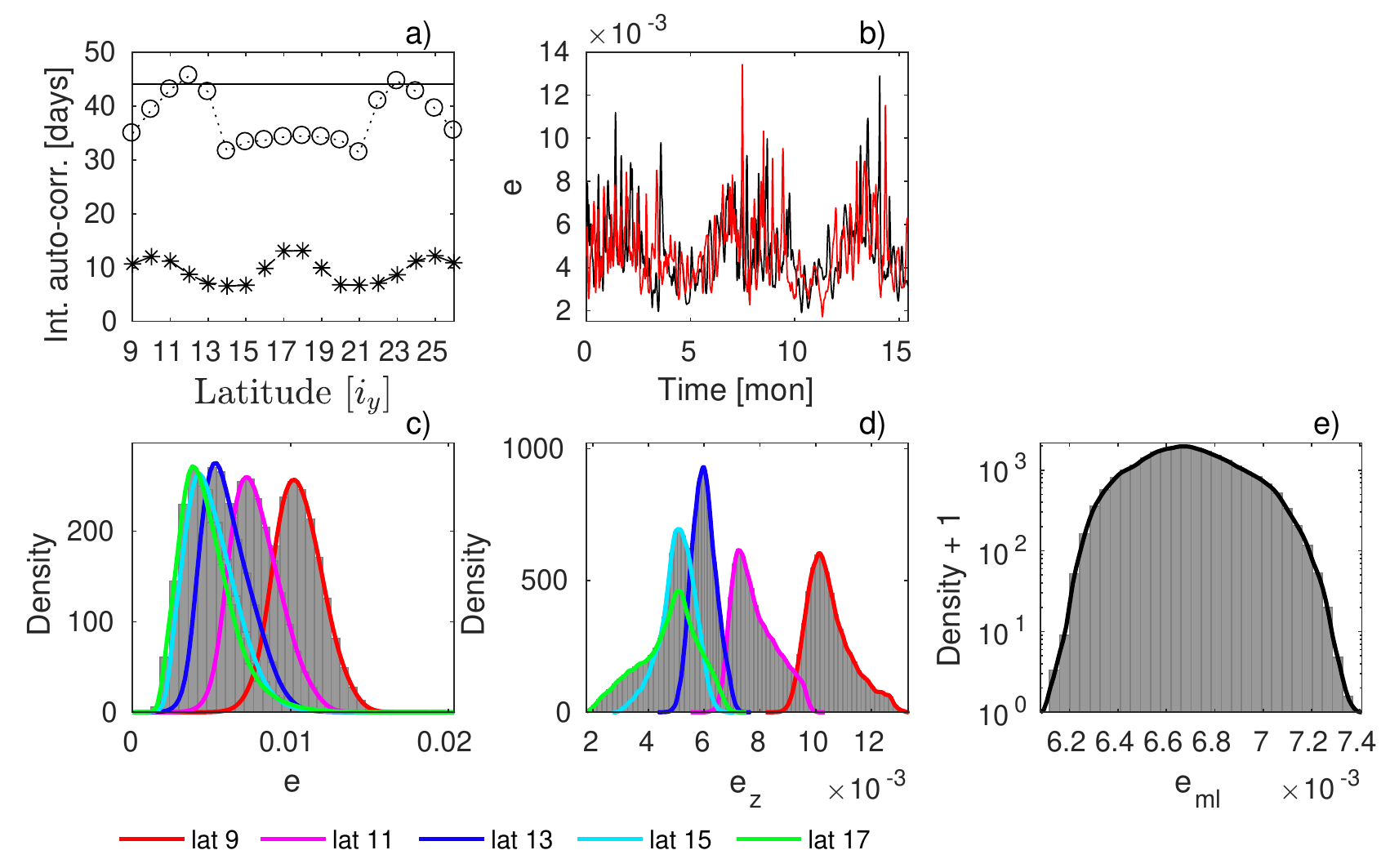}
  \caption{Statistical properties of the total energy for $\Delta T=40$ K. (a) Integrated auto-correlation time scales for the local (dotted lines with star markers), zonally-averaged (dashed lines with circle markers), and average mid-latitude (continuous line) observables. (b) Time series of the local energy at latitude $i_y=17$ and at two different longitudes: $i_x=4$ (red line) and $i_x=19$ (black line). Histograms of the (c) local, (d) zonally-averaged, and (e) average mid-latitude observables. In the case of (c) -- (e), the continuous lines show the approximation of the PDFs by kernel smoothing (\textit{ksdensity} function in MATLAB); the colours mark different latitudes according to the legend.} 
  \label{fig:stat6}
\end{figure}

For the analysis of extreme events, we use a similar procedure as in the case of strong forcing ($\Delta T=133$ K), and concatenate the three-monthly maxima series for every second longitude one after the other in $x$-direction. Thus, we increase the length of available data for the local observables to about $8.7 \times 10^6$ (from about $4.8 \times 10^5$) years. Although the time series at every second longitude are correlated with each other, the correlation almost vanishes at extreme levels, being below 0.1 for every latitude in the case of the 8-years BM. We define the GP exceedance ratios so that the number of threshold exceedances corresponds to the number of block maxima, as described in Sec.~\ref{sec:tg20}.

In the case of weak forcing, the theoretical shape parameter is -0.03, shown by the grey horizontal line in Fig.~\ref{fig:shape6}. The grey shading represents the range of theoretical values resulting from taking into consideration possible geometrical degeneracies according to the limits described in Sec. \ref{sec:theory}. We plot the GEV shape parameter against exponentially increasing block sizes of $n=2^{i}$ years, where $i=0,...,15$ for the local observables and $i=0,...,11$ for the zonally-averaged and average mid-latitude observables. Focusing first on the local observables, we notice a non-monotonic change of the shape parameter according to increasing block sizes. For block sizes smaller than 30 years, the shape parameter even reaches non-physical, positive values for certain latitudes. \textcolor{black}{This change of sign of the estimated shape parameters is similar to what has been observed by \citet{vannitsem2007} in case of local temperature extremes in a more realistic QG model with orography.} The non-monotonic changes and the positive shape parameter estimates have to do with the fact that, if the block size is not large enough, we select events from both regimes (more and less fluctuating) thus ``contaminating'' the statistics of extremes; whereas if the block size is large enough, only extremes from the more unstable regime are selected. Figure~\ref{fig:shape6} (a) also shows that the estimated shape parameter seems to converge at almost every latitude to a value which is lower than the theoretical shape parameter, yet near to the range of values obtained taking into consideration possible geometrical degeneracies, see Sec. \ref{sec:theory}. As discussed above, this is in fact unsurprising given the qualitative properties of the system in the low forcing regime.

In case of the zonally-averaged and average mid-latitude observables we cannot detect any convergence. This is an expected result, considering the statistical and dynamical characteristics of our data and the fact that the length of the time series is in this case even shorter than for the local observables. As an effect of the ``shoulders'' in the PDFs, we obtain very uncertain estimates even for large block sizes, and the KS-tests reject the hypothesis of a GEV model in these cases. The shape parameter estimates have a large latitudinal spread due to the varying form of PDFs according to latitudes. Except the differences between the GEV (Fig.~\ref{fig:shape6}, a~--~c) and GP (\ref{fig:shape6}, d~--~f) shape parameters at small block sizes and high exceedance ratios , both methods show us basically the same picture. The misleading property of the KS-test p-values $p$ is underlined by Fig.~\ref{fig:shape6}. Even in case of the zonally-averaged and average mid-latitude observables, where we cannot detect any convergence at all, we find $p>0.05$ for a wide range of block sizes and exceedance ratios (circle markers).

\begin{figure}[ht!]
  \centering
  \includegraphics[width=1\textwidth]{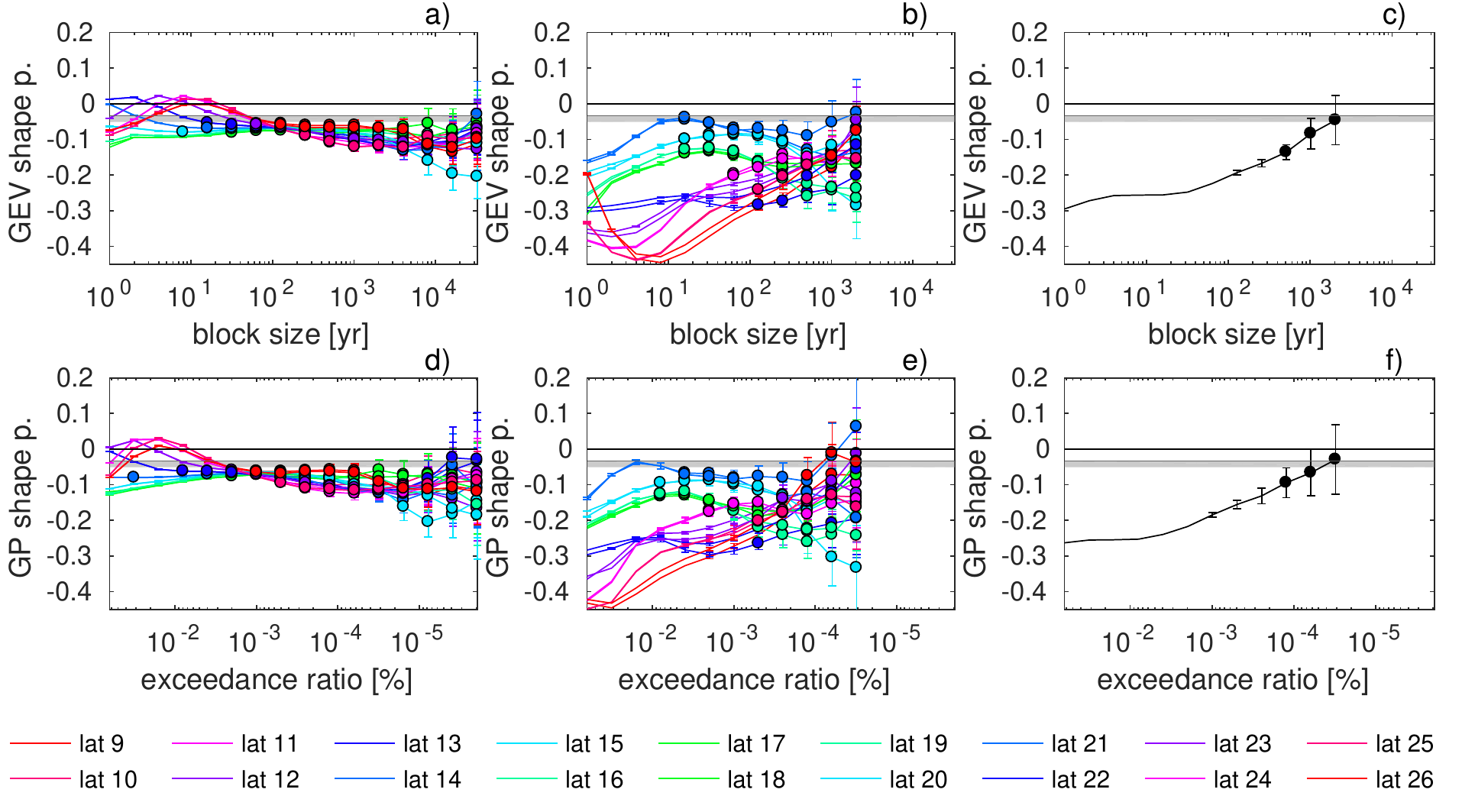}
  \caption{Shape parameter for $\Delta T=40$ K. GEV shape parameter for (a) local, (b) zonally-averaged, and (c) average mid-latitude energy. GP shape parameter for (d) local, (e) zonally-averaged, and (f) average mid-latitude energy. The circle markers point out shape parameter values, for which the p-value of the KS-test is above 0.05. Grey, horizontal line: theoretical shape parameter. The grey shading represents a possible range of the theoretical shape parameter according to the limits described in Sec. \ref{sec:theory}. The error bars show 95~\% confidence intervals of the maximum likelihood estimation. Different colours represent different latitudes} 
  \label{fig:shape6}
\end{figure}

 \section{Summary and discussion}
 
In this paper we have studied the convergence of statistically estimated GEV and GP shape parameters to a a theoretical shape parameter. The latter is calculated based on properties of the attractor \citep{holland2012,lucarini2014}. We analyse a quasi-geostrophic 2-layer atmospheric model. We study extremes of different types of energy observables: local, zonally-averaged, and average mid-latitude energy. We perform simulations with two different forcing levels: a strong forcing ($\Delta T=133$ K), producing a highly chaotic behaviour of the system, and a weak forcing ($\Delta T=40$ K), producing a less pronounced chaotic behaviour. In the case of strong (weak) forcing, we produce time series of about $9.6 \times 10^4$ ($4.8 \times 10^5$) years, representing a deterministic equivalent to a stationary process. We estimate the GEV and GP shape parameters for exponentially increasing block sizes and exponentially decreasing exceedance ratios (fractions of above-threshold events), i.e., increasing thresholds, by performing maximum likelihood estimation. For comparability, we choose the GP thresholds so that the number of threshold exceedances corresponds to the number of block maxima. We take advantage of the fact that statistics are uniform in the zonal direction, and use the data from every second longitude for the analysis of extreme events, thus increasing the length of available data for the local observables to about $1.7 \times 10^6$ ($8.7 \times 10^6$) years in the case of strong (weak) forcing.

We start the discussion of our results with the strong forcing regime. In this case, we observe a roughly monotonic increase of the estimated GEV (GP) shape parameters towards the theoretical value $\xi_\delta=-0.002$. The estimated shape parameters seem to converge to $\xi_\delta$ in case of the local observables at every latitude and in case of the zonally-averaged observables at central latitudes. Thus, our numerical results allow for robust conclusions regarding the universality of extremes, according to the theory presented in Sec.~\ref{sec:theory}. However, in the case of several (especially marginal) latitudes of the zonally-averaged observables, as well as for the average mid-latitude observable, the estimated shape parameter is relatively far from the theoretical one. For these observables the amount of data seems to be not enough to approach asymptotic levels, thus we cannot make more precise statements on the success of the analysis. Even in this extremely chaotic case, the convergence is very slow, suggesting that customary choices like yearly maxima are not always the best option for an accurate modelling of extremes.

Despite the predicted universal asymptotic properties of extremes, if we consider a certain block size (threshold), we find that the shape parameter estimates are different among the observables and latitudes. Thus, on pre-asymptotic level, extremes show rather diverse properties. The speed of convergence to the asymptotic level is not universal. The local observables exhibit high-frequency fluctuations, as an effect of boundary fluxes, and at the same time, the fastest convergence of the shape parameter estimates to the theoretical value. Since the energy is transported mostly along the zonal direction by the zonal mean flow, by averaging along a latitudinal band the highest frequencies are filtered out, and fluctuations with lower frequencies become stronger. In the case of the zonally-averaged observables, we obtain weak auto-correlations and fast convergence to $\xi_\delta$ in the middle of the channel where the baroclinicity is the strongest, versus high auto-correlations and slow convergence at the margins of the channel where instead the baroclinicity is weak. The stronger the serial correlation, the less the number of uncorrelated data in a block, and the larger block sizes are needed in order to approach asymptotic levels (see also \citep{coles2001}). By averaging over the mid-latitude area, one merges zonally-averaged time series exhibiting different auto-correlations. Thereby, the convergence to $\xi_\delta$ is faster than in case of the zonally-averaged observables at marginal latitudes. To sum up, a very important conclusion of our study is the existence of latitude-dependent pre-asymptotic differences, as a counterpart to the universal asymptotic properties.

We assume that the extremely slow convergence has to do mainly with the fact that $\xi_\delta$ is negative but very close to 0. Based on $\xi_\delta$ and on the estimated GP modified scale parameter, one is able to estimate according to \citet{lucarini2014} the absolute maximum, which is the upper end point of the GP distribution, as mentioned in Sec.~\ref{sec:tg20}. By performing a very rough estimation (and neglecting the weak latitude-dependence of the GP modified scale parameter), the absolute maximum in case of the local observables $A_{max}\approx 12.5$, which is about $200$ times the mean local energy value (see Fig. \ref{fig:stat}) and 20 times larger than some of the largest estimated return levels obtained for the largest return times considered here (see Fig.~\ref{fig:rl20}). This means that extremes are bounded, and an absolute maximum does exist, but the tail is extremely stretched out, and ultra long simulations are needed to explore this absolute maximum. Our results point out the discrepancy between the existence of a mathematical limit and the actual possibility of observing it. Note that if the asymptotic shape parameter is lower, the absolute maximum will be much closer to the maximum observed within a long yet finite time series, as it is shown in a recent study on temperature extremes in Southern Pakistan \citep{zahidetal2017}. 

Our conclusions regarding the convergence of the estimated shape parameter to $\xi_\delta$ are confirmed by results based on the GP modified scale and return level estimates, in the case of the local observables. We point out, however, that the longer the return period, the slower the convergence of the estimated return levels to their asymptotic values, and the larger the underestimation of the asymptotic return levels if we consider small block sizes (low thresholds).
 
In the case of weak forcing, temporal and spatial correlations are very strong due to a regime behaviour of our system, which exhibits two regimes: a more unstable one with stronger fluctuations and a less unstable one with reduced fluctuations. Due to this regime behaviour the statistics of extreme events are ``contaminated'': if the block size (threshold) is not large (high) enough, we select events from both regimes, whereas if it is large (high) enough, only extremes from the more unstable regime are selected. This induces non-monotonic changes of the estimated shape parameters  by increasing the block size (threshold), and leads to the appearance of positive, i.e., non-physical, or very low shape parameter estimates. In the case of the local observables, the estimated shape parameters seem to converge at almost every latitude to a value which is lower ($\approx -0.06$) than the theoretical shape parameter ($\xi_\delta=-0.03$). Furthermore, in the case of the zonally-averaged and average mid-latitude observables, we cannot detect any convergence at all. The inconsistency of our numerical results with the theory is, in fact, unsurprising given the qualitative properties of the system in the low forcing regime, which do not resemble characteristics of Axiom A systems (at least on the finite time scales we are able to explore based on the available data).

Our results show that with increasing block size or threshold the shape parameters of the GEV and GP distributions are becoming more and more similar, according to the asymptotic equivalence of the two models \citep{coles2001, lucarini2014}. Both methods show us basically the same picture regarding the statistical properties of extreme events. Despite the mentioned similarities, we observe also some differences between the two approaches. The convergence to the limiting distribution seems to be somewhat faster in the case of the POT approach. This is in agreement with the well-established fact that the POT approach produces often more accurate predictions in case of applications \citep{davison1990, coles2001}. Despite the faster convergence, however, the best GP shape parameter estimates (defined in Sec.~\ref{sec:tg20}) do not approximate $\xi_\delta$ more accurately than the best GEV shape parameter estimates. Therefore, the advantage of the POT approach compared the BM approach is irrelevant in the case of very long time series.
 
We use the Kolmogorov-Smirnov test (KS-test) to verify the fit of the GEV (GP) distribution to the distribution of extremes, selected as block maxima (threshold exceedances). Our results show that the KS-test is merely an indicator of the fit quality, and does not show whether the convergence to the correct GEV (GP) distribution is reached or not. The KS test suggests a good fit to the GEV (GP) distribution even in cases when the distance between the estimated and the asymptotic shape parameter is substantial and even if no convergence can be detected. The misleading property of p-values of the KS and Pearson's chi-squared tests was also pointed out in previous studies in the case of more simple systems \citep{faranda2011,bodai2017}. In this work, we estimate the GEV and GP parameters performing maximum likelihood estimation \citep{coles2001}, but it would be relevant to find out to what extent other estimation procedures, like the L-moments \citep{hosking1990} or probability-weighted moments methods \citep{hoskingetal1985}, would change the results.
 
Concluding, we would like to emphasise some key messages one can get from our results: 
\begin{itemize}	
\item Indeed, we have been able to find the signature of the universal properties of the extremes of physical observables in strongly chaotic dynamical systems, as predicted in the case of Axiom A systems. Nonetheless, given the availability of very long yet finite time series, we have been able to find more convincing results (yet with a relatively large uncertainty) only for specific observables, because in the case of observables featuring serial correlations it is extremely hard to collect robust statistics of extremes. 
\item We have observed that in the case of strong forcing the estimate of the shape parameter increases monotonically towards its asymptotic value for stricter and stricter criteria of selection of extremes. This corresponds to the fact that we manage to collect more detailed information on the local properties of the attractor near the point of absolute maximum of the observable, and thus explore all the dimensions of the attractor.
\item We also remark that agreement of the results with the theory of extremes of observables of dynamical systems developed in the context of Axiom A flows cannot be found in the case of the weakly chaotic flow featuring regime behaviour and strong spatial and temporal correlations, according to what one would expect considering the chaotic hypothesis. 
\item Finally, we note that the predicted and estimated shape parameters are extremely small so that the statistics of extremes is virtually indistinguishable, up to ultra long return periods, from what would be predicted by a Gumbel distribution ($\xi=0$), which emerges as the statistical model of reference for physical extremes in high dimensional chaotic systems, and suggests in the case of fluids the existence of a well-developed turbulent state. \textcolor{black}{However, if the considered observable lives on a lower dimensional attractor, it is theoretically possible that the asymptotic shape parameter is much lower, and thus distinguishable from 0. In case of applications, one deals with data sets having heterogeneous properties (non-smooth observables, existence of multiple time scales, effects of spatial inhomogenities), and thus the estimated shape parameter can exhibit change of signs, fluctuations, or (maybe even several) stable values only for a limited range of block sizes or thresholds (some of these features can be observed in our study in case of weak forcing, but can be found also in lower dimensional models, as shown by Bodai 2017, or even in more realistic models with orography, as observed by Vannitsem 2007). In some cases, an extreme value law does not even exist or is extremely difficult to be explored, and in other cases other distributions than the Gumbel, i.e. Frechet or Weibull, can be more appropriate for the estimation of return levels up to a limited return period.}
\end{itemize}

\section*{Acknowledgements}

We would like to thank to Sebastian Schubert, Christian Franzke, Maida Zahid and Richard Blender for useful discussions. We are indebted to Sebastian Schubert for his support in performing the simulations and providing the code for computing the Lyapunov Exponents and Kaplan-York dimensions. Valerio Lucarini acknowledges the many exchanges on these topics with Davide Faranda, Antonio Speranza, and Renato Vitolo. Valerio Lucarini also acknowledges support received from the Sfb/Transregio project TRR181. V. M. G\'alfi acknowledges funding from the International Max Planck Research School on Earth System Modelling (IMPRS-ESM) The authors acknowledge an anonymous referee for suggesting an alternative definition of the accuracy.

\section*{Conflicts of interest}

The author(s) declare(s) that there is no conflict of interest regarding the publication of this paper.

\newpage
\pagenumbering{Roman}						
\setcounter{page}{1}

\renewcommand\bibname{References}



\end{document}